\documentclass[10pt,a4paper]{article}
\usepackage[utf8]{inputenc}
\usepackage[T1]{fontenc}
\usepackage[english]{babel}
\usepackage{hyperref}
\usepackage{amsmath,amsthm,enumerate}
\usepackage{amssymb}
\usepackage{mathrsfs}
\usepackage{tikz}\usetikzlibrary{snakes}
\usepackage{subfigure}
\usepackage[hmargin=2.5cm,vmargin=3cm]{geometry}
\usepackage[color=green!30]{todonotes}
\usepackage{bm}

\newtheorem{thm}{Theorem}

\newtheorem{prop}[thm]{Proposition}
\newtheorem{lem}[thm]{Lemma}

\newtheorem{cor}[thm]{Corollary}

\newtheorem{defi}[thm]{Definition}
\newtheorem{prope}[thm]{Property}

\newtheorem{claimbis}[thm]{Claim}

\newcommand{\decisionpb}[4]{
        \begin{minipage}{#4\textwidth}
                #1\\
                \emph{Instance:} #2\\ 
                \emph{Question:} #3
        \end{minipage}
}

\newcommand{\ID}{\gamma^{\text{\tiny{ID}}}}
\newcommand{\LD}{\gamma^{\text{\tiny{LD}}}}
\newcommand{\OLD}{\gamma^{\text{\tiny{OLD}}}}

\newcommand{\MD}{dim}

\newcommand{\PBID}{\textsc{Identifying Code}}
\newcommand{\PBLD}{\textsc{Locating-Dominating-Set}}
\newcommand{\PBOLD}{\textsc{Open Locating-Dominating Set}}
\newcommand{\PBMD}{\textsc{Metric Dimension}}

\newcommand{\mS}{\mathsf{S}}
\newcommand{\msep}{\mathsf{sep}}
\newcommand{\msepr}{\mathsf{toSepR}}
\newcommand{\mcnt}{\mathsf{cnt}}

\newenvironment{proofofclaim}{\noindent \textit{Proof of claim. }}{\hfill{\tiny($\Diamond$)}\medskip}

\begin{document}

\title{Identification, location-domination and metric dimension on interval and permutation graphs. II. Algorithms and complexity\footnote{A short version of this paper, containing only the results about location-domination and metric dimension, appeared in the proceedings of the WG 2015 conference~\cite{WGversion}.}}
\author{Florent Foucaud\footnote{\noindent LIMOS, Universit\'e Blaise Pascal, Clermont-Ferrand (France). florent.foucaud@gmail.com}
\and George B. Mertzios\footnote{\noindent School of Engineering and Computing Sciences, Durham University (United Kingdom). george.mertzios@durham.ac.uk} 
\footnote{\noindent Partially supported by the EPSRC Grant EP/K022660/1.}
\and Reza Naserasr\footnote{\noindent CNRS - IRIF, Universit\'e Paris-Diderot, Paris (France). reza@irif.fr}
\and Aline Parreau\footnote{\noindent CNRS - LIRIS, Université Lyon 1, Villeurbanne (France). aline.parreau@univ-lyon1.fr}
\and Petru Valicov\footnote{\noindent LIF, Université d'Aix-Marseille, Marseille (France). petru.valicov@lif.univ-mrs.fr}}

\maketitle

\begin{abstract}
We consider the problems of finding optimal identifying codes, (open) locating-dominating sets and resolving sets (denoted \PBID, \textsc{(Open) Locating-Dominating Set} and \PBMD) of an interval or a permutation graph. In these problems, one asks to distinguish all vertices of a graph by a subset of the vertices, using either the neighbourhood within the solution set or the distances to the solution vertices. Using a general reduction for this class of problems, we prove that the decision problems associated to these four notions are NP-complete, even for interval graphs of diameter~$2$ and permutation graphs of diameter~$2$. While \PBID{} and \textsc{(Open) Locating-Dominating Set} are trivially fixed-parameter-tractable when parameterized by solution size, it is known that in the same setting \PBMD{} is $W[2]$-hard. We show that for interval graphs, this parameterization of \PBMD{} is fixed-parameter-tractable.
\end{abstract}

\section{Introduction}

Combinatorial identification problems have been widely studied in various contexts. The common characteristic of these problems is that we are given a combinatorial structure (graph or hypergraph), and we wish to distinguish (i.e. uniquely identify) its vertices by the means of a small set of selected elements. In this paper, we study several such related identification problems where the instances are graphs. In the problem \emph{metric dimension}, we wish to select a set $S$ of vertices of a graph $G$ such that every vertex of $G$ is uniquely identified by its distances to the vertices of $S$. The notions of \emph{identifying codes} and \emph{(open) locating-dominating sets} are similar. Roughly speaking, instead of the distances to $S$, we ask for the vertices to be distinguished by their neighbourhood within $S$. These problems have been widely studied since their introduction in the 1970s and 1980s. They have been applied in various areas such as network verification~\cite{BBDGKP11,BEEHHMR06}, fault-detection in networks~\cite{KCL98,UTS04}, graph isomorphism testing~\cite{B80} or the logical definability of graphs~\cite{KPSV04}. We note that the similar problem of finding a \emph{test cover} of a hypergraph (where hyperedges distinguish the vertices) has been studied under several names by various authors, see e.g.~\cite{BS07,B72,CCCHL08,GJ79,MS85,R61}.\\[-2mm]

\noindent\textbf{Important concepts and definitions.} All considered graphs are finite and simple. We will denote by $N[v]$, the \emph{closed neighbourhood} of vertex $v$, and by $N(v)$ its \emph{open neighbourhood}, i.e. $N[v]\setminus\{v\}$. A vertex is \emph{universal} if it is adjacent to all the vertices of the graph. A set $S$ of vertices of $G$ is a \emph{dominating set} if for every vertex $v$, there is a vertex $x$ in $S\cap N[v]$. It is a \emph{total dominating set} if instead, $x\in S\cap N(v)$. In the context of (total) dominating sets we say that a vertex $x$ \emph{(totally) separates} two distinct vertices $u,v$ if it (totally) dominates exactly one of them.
Set $S$ (totally) separates the vertices of a set $X$ if every pair in $X$ has a vertex in $S$ (totally) separating it. We have the three key definitions, that merge the concepts of (total) domination and (total) separation:

\begin{defi}[Slater~\cite{Sl87,S88}, Babai~\cite{B80}] A set $S$ of vertices of a graph $G$ is a \emph{locating-dominating set} if it is a dominating set and it separates the vertices of $V(G)\setminus S$.
\end{defi}

The smallest size of a locating-dominating set of $G$ is the \emph{location-domination number} of $G$, denoted $\LD(G)$. This concept has also been used under the name \emph{distinguishing set} in~\cite{B80} and \emph{sieve} in~\cite{KPSV04}.

\begin{defi}[Karpovsky, Chakrabarty and Levitin~\cite{KCL98}] A set $S$ of vertices of a graph $G$ is an \emph{identifying code} if it is a dominating set and it separates all vertices of $V(G)$.
\end{defi}

The smallest size of an identifying code of $G$ is the \emph{identifying code number} of $G$, denoted $\ID(G)$.

\begin{defi}[Seo and Slater~\cite{SS10}] A set $S$ of vertices of a graph $G$ is an \emph{open locating-dominating set} if it is a total dominating set and it totally separates all vertices of $V(G)$.
\end{defi}

The smallest size of an open locating-dominating set of $G$ is the \emph{open location-domination number} of $G$, denoted $\OLD(G)$. This concept has also been called \emph{identifying open code} in~\cite{HY12}.

Another kind of separation based on distances is used in the following concept:

\begin{defi}[Harary and Melter~\cite{HM76}, Slater~\cite{S75}] A set $R$ of vertices of a graph $G$ is a \emph{resolving set} if for each pair $u,v$ of distinct vertices, there is a vertex $x$ of $R$ with $d(x,u)\neq d(x,v)$.\footnote{Resolving sets are also known under the name of \emph{locating sets}~\cite{S75}. Optimal resolving sets have sometimes been called \emph{metric bases} in the literature; to avoid an inflation in the terminology we will only use the term \emph{resolving set}.}
\end{defi}

The smallest size of a resolving set of $G$ is the \emph{metric dimension} of $G$, denoted $\MD(G)$. 

It is easy to check that the inequalities $\MD(G)\leq \LD(G)\leq \ID(G)$ and $\LD(G)\leq \OLD(G)$ hold, indeed every locating-dominating set of $G$ is a resolving set, and every identifying code (or open locating-dominating set) is a locating-dominating set. Moreover it is proved that $\ID(G)\leq 2\LD(G)$~\cite{GKM08} (using the same proof idea one would get a similar relation between $\LD(G)$ and $\OLD(G)$ and between $\ID(G)$ and $\OLD(G)$, perhaps with a different constant). There is no strict relation between $\ID(G)$ and $\OLD(G)$.

In a graph $G$ of diameter~$2$, one can easily see that the concepts of resolving set and locating-dominating set are almost the same, as $\LD(G)\leq\MD(G)+1$. Indeed, let $S$ be a resolving set of $G$. Then all vertices in $V(G)\setminus S$ have a distinct neighbourhood within $S$. There might be (at most) one vertex that is not dominated by $S$, in which case adding it to $S$ yields a locating-dominating set.

While a resolving set and a locating-dominating set exist in every graph $G$ (for example the whole vertex set), an identifying code may not exist in $G$ if it contains \emph{twins}, that is, two vertices with the same closed neighbourhood. However, if the graph is \emph{twin-free} the set $V(G)$ is an identifying code of $G$. Similarly, a graph admits an open locating-dominating set if and only if it has no \emph{open twins}, i.e. vertices sharing the same open neighbourhood. We say that such a graph is \emph{open twin-free}.

The focus of this paper is the following set of four decision problems:
\medskip

\noindent\decisionpb{\PBLD}{A graph $G$, an integer $k$.}{Is it true that $\LD(G)\leq k$?}{0.5}
\noindent\decisionpb{\PBID}{A graph $G$, an integer $k$.}{Is it true that $\ID(G)\leq k$?}{0.5}\\[0.5em]

\noindent\decisionpb{\PBOLD}{A graph $G$, an integer $k$.}{Is it true that $\OLD(G)\leq k$?}{0.5}
\noindent\decisionpb{\PBMD}{A graph $G$, an integer $k$.}{Is it true that $\MD(G)\leq k$?}{0.5}
\medskip

We will study these four concepts and decision problems on graphs belonging to specific subclasses of perfect graphs (i.e. graphs whose induced subgraphs all have equal clique and chromatic numbers). Many standard graph classes are perfect, for example bipartite graphs, split graphs, interval graphs. For precise definitions, we refer to the books of Brandstädt, Le and Spinrad and of Golumbic~\cite{BL99,golumbicBook}. Some of these classes are classes defined using a geometric intersection model, that is, the vertices are associated to the elements of a set $S$ of (geometric) objects, and two vertices are adjacent if and only if the corresponding objects intersect. The graph defined by the intersection model $S$ is its \emph{intersection graph}. An \emph{interval graph} is the intersection graph of intervals of the real line, and a \emph{unit interval graph} is an interval graph whose intersection model contains only (open) intervals of unit length. Given two parallel lines $B$ and $T$, a \emph{permutation graph} is the intersection graph of segments of the plane which have one endpoint on $B$ and the other endpoint on $T$.

Interval graphs and permutation graphs are classic graph classes that have many applications and are widely studied. They can be recognized efficiently, and many combinatorial problems have simple and efficient algorithms for these classes.\\[-2mm]

\noindent\textbf{Previous work.} The complexity of distinguishing problems has been studied by many authors. \PBID{} was first proved to be NP-complete by Charon, Hudry, Lobstein and Z\'emor~\cite{CHLZ01}, and \PBLD{}, by Colbourn, Slater and Stewart~\cite{CSS87}. Regarding their instance restriction to specific graph classes, \PBID{} and \PBLD{} were shown to be NP-complete for bipartite graphs by Charon, Hudry and Lobstein~\cite{CHL03}. This was improved by M\"uller and Sereni to planar bipartite unit disk graphs~\cite{MS09}, by Auger to planar graphs with arbitrarily large girth~\cite{A10}, and by Foucaud to planar bipartite subcubic graphs~\cite{F13j}. Foucaud, Gravier, Naserasr, Parreau and Valicov proved that \PBID{} is NP-complete for graphs that are both planar and line graphs of subcubic bipartite graphs~\cite{lineID}. Berger-Wolf, Laifenfeld, Trachtenberg~\cite{BWLT06} and Suomela~\cite{SUOM07} independently showed that both \PBID{} and \PBLD{} are hard to approximate within factor $\alpha$ for any $\alpha=o(\log n)$ (where $n$ denotes the order of the graph), with no restriction on the input graph. This result was recently extended to bipartite graphs, split graphs and co-bipartite graphs by Foucaud~\cite{F13j}. Moreover, Bousquet, Lagoutte, Li, Parreau and Thomassé \cite{BLLPT14} proved the same non-approximability result for bipartite graphs with no 4-cycles. On the positive side, \PBID{} and \PBLD{} are constant-factor approximable for bounded degree graphs (showed by Gravier, Klasing and Moncel in \cite{GKM08}), line graphs~\cite{F13j,lineID}, interval graphs~\cite{BLLPT14} and are linear-time solvable for graphs of bounded clique-width (using Courcelle's theorem~\cite{C90}). Furthermore, Slater~\cite{Sl87} and Auger~\cite{A10} gave explicit linear-time algorithms solving \PBLD{} and \PBID, respectively, in trees.

The complexity of \PBOLD{} was not studied much; Seo and Slater showed that it is NP-complete~\cite{SS10}, and the inapproximability results of Foucaud~\cite{F13j} for bipartite, co-bipartite and split graphs transfer to it.

The problem \PBMD{} is widely studied. It was shown to be NP-complete by Garey and Johnson~\cite[Problem GT61]{GJ79}. This result has recently been extended to bipartite graphs, co-bipartite graphs, split graphs and line graphs of bipartite graphs by Epstein, Levin and Woeginger~\cite{ELW12j}, to a special subclass of unit disk graphs by Hoffmann and Wanke~\cite{HW12}, and to planar graphs by Diaz, Pottonen, Serna and van Leeuwen~\cite{DPSV12}.

Epstein, Levin and Woeginger~\cite{ELW12j} also gave polynomial-time algorithms for the weighted version of \PBMD{} for paths, cycles, trees, graphs of bounded cyclomatic number, cographs and partial wheels. Diaz, Pottonen, Serna, van Leeuwen~\cite{DPSV12} gave a polynomial-time algorithm for outerplanar graphs, and Fernau, Heggernes, van't Hof, Meister and Saei gave one for chain graphs~\cite{FHvMS15}. \PBMD{} can most likely not be expressed in MSOL and it is an open problem to determine its complexity for bounded treewidth (even treewidth 2).

\PBMD{} is hard to approximate within any $o(\log n)$ factor for general graphs, as shown by Beerliova, Eberhard, Erlebach, Hall, Hoffmann, Mihal\'ak and Ram~\cite{BEEHHMR06}. This is even true for subcubic graphs, as shown by Hartung and Nichterlein~\cite{HN12} (a result extended to \emph{bipartite} subcubic graphs in Hartung's thesis~\cite{seppThesis}).

In light of these results, the complexity of \PBLD, \PBOLD, \PBID{} and \PBMD{} for interval and permutation graphs is a natural open question (as asked by Manuel, Rajan, Rajasingh, Chris Monica M.~\cite{honeycomb} and by Epstein, Levin, Woeginger~\cite{ELW12j} for \PBMD{} on interval graphs), since these classes are standard candidates for designing efficient algorithms to solve otherwise hard problems.

Let us say a few words about the parameterized complexity of these problems. A decision problem is said to be fixed-parameter tractable (FPT) with respect to a parameter $k$ of the instance, if it can be solved in time $f(k)n^{O(1)}$ for an instance of size $n$, where $f$ is a computable function (for definitions and concepts in parameterized complexity, we refer to the books~\cite{DF13,N06}). It is known that for the problems \PBLD, \PBOLD{} and \PBID, for a graph of order~$n$ and solution size~$k$, the bound $n\leq 2^k$ holds (see e.g.~\cite{KCL98,S88}). Therefore, when parameterized by~$k$, these problems are (trivially) FPT: one can first check whether $n\leq 2^k$ holds (if not, return ``no''), and if yes, use a brute-force algorithm checking all possible subsets of vertices. This is an FPT algorithm. However, \PBMD{} (parameterized by solution size) is W[2]-hard even for bipartite subcubic graphs~\cite{seppThesis,HN12} (and hence unlikely to be FPT). Remarkably, the bound $n\leq D^k+k$ holds~\cite{CEJO00} (where $n$ is the graph's order, $D$ its diameter, and $k$ is the solution size of \PBMD), and therefore for graphs of diameter bounded by a function of~$k$, the same arguments as the previous ones yield an FPT algorithm. This holds, for example, for the class of (connected) split graphs, which have diameter at most~$3$. Also, it was recently proved that \PBMD{} is FPT when parameterized by the largest possible number of leaves in a spanning tree of a graph~\cite{E15}. Besides this, as remarked in~\cite{HN12}, no non-trivial FPT algorithm for \PBMD{} was previously known.

Finally, we also mention a companion paper~\cite{part1}, in which we study problems \PBID, \PBLD, \PBOLD{} and \PBMD{} on interval and permutation graphs from a combinatorial point of view, proving several bounds involving the order, the diameter and the solution size of a graph.\\[-2mm]

\noindent\textbf{Our results.} We continue the classification of the complexity of problems \PBID, \PBLD, \PBOLD{} and \PBMD{} by giving a unified reduction showing that all four problems are NP-complete even for graphs that are interval graphs and have diameter~$2$ or permutation graphs of diameter~$2$. The reductions are presented in Section~\ref{sec:reduction}. Then, in Section~\ref{sec:MD-interval-FPT}, we use dynamic programming on a path-decomposition to show that \PBMD{} is FPT on interval graphs, when the parameter is the solution size. Up to our knowledge, this is the first non-trivial FPT algorithm for this problem when parameterized by solution size. We then conclude the paper with some remarks in Section~\ref{sec:conclu}.

\section{Hardness results}\label{sec:reduction}

We will provide a general framework to prove NP-hardness for distinguishing problems in interval graphs and permutation graphs. We just need to assume few generic properties about the problems, and then provide specific gadgets for each problem.

We will reduce our problems from \textsc{3-Dimensional Matching} which is known to be NP-complete~\cite{K72}.

\medskip

\noindent\decisionpb{{\sc 3-Dimensional Matching}}{Three disjoint sets $A$, $B$ and $C$ each of size $n$, and a set $\mathcal T$ of $m$ triples of $A\times B\times C$.}
{Is there a perfect 3-dimensional matching $\mathcal M\subseteq \mathcal T$ of the hypergraph $(A, B, C,\mathcal T)$, i.e. a set of disjoint triples of $\mathcal T$ such that each element of $A\cup B\cup C$ belongs to exactly one of the triples?}{}

We give the general framework and the gadgets we will use in Section~\ref{sec:prel}, then prove the general reduction using this framework in Section~\ref{sec:reduc} and apply it to obtain the NP-hardness for \PBID, \PBLD{} and \PBOLD{} in Section~\ref{sec:app}. We finally deduce from the results for graphs of diameter~$2$ the hardness of \PBMD{} in Section~\ref{sec:diam2}. We give the reduction using interval graphs and prove subsequently that it can be built as a permutation graph too.

\subsection{Preliminaries and gadgets}\label{sec:prel}

In the three distinguishing problems \PBLD, \PBID{} and \PBOLD, one
asks for a set of vertices that dominates all vertices and separates
all pairs of vertices (for suitable definitions of domination and
separation). Since we give a reduction which applies to all three
problems (and others that share certain properties with them), we will
generally speak of a \emph{solution} as a vertex set satisfying the
two properties.

For two vertices $u,v$ let us denote by $I_{u,v}$ the set
$N[u]\setminus N[v]$. In the reduction, we will only make use of the
following properties (that are common to \PBLD, \PBID{} and \PBOLD):

\begin{prope}\label{prop:domsep}
 Let $G$ be a graph with a solution $S$ to \PBLD, \PBID{} or \PBOLD.\\
(1) For each vertex $v$, any vertex from $N(v)$ dominates $v$;\\
(2) For each vertex $v$, at least one element
of $N[v]$ belongs to $S$;\\
(3) For every pair $u,v$ of adjacent vertices, any vertex of $I_{u,v}\cup I_{v,u}$ separates $u,v$;\\
(4) For every pair $u,v$ of adjacent vertices,
$S$ contains a vertex of $I_{u,v}\cup I_{v,u}\cup \{u,v\}$.
\end{prope}

The problems \PBID{} and \PBOLD{} clearly satisfy these
properties. For \PBLD, the vertices of a solution set $S$ do not need
to be separated from any other vertex. However one can say
equivalently that two vertices $u,v$ are separated if either $u$ or
$v$ belongs to $S$, or if there is a vertex of $S$ in $I_{u,v}\cup
I_{v,u}$. Therefore, \PBLD{} also satisfies the above properties.

Before describing the reduction, we define the following
\emph{dominating gadget} independently of the considered problem (we
describe the specific gadgets for \PBLD, \PBOLD{} and \PBID{} in
Section~\ref{sec:app}). The idea behind this gadget is to
ensure that specific vertices are dominated locally --- and are therefore
separated from the rest of the graph. We will use it extensively in
our construction.

\begin{defi}[Dominating gadget]\label{def:domgadget}
A \emph{dominating gadget} $D$ is an interval graph such that there
exists an integer $d\geq 1$ and a subset $S_D$ of $V(D)$ of size $d$
(called \emph{standard solution for $D$}) with the following
properties:
\begin{itemize}
\item $S_D$ is an optimal solution for $D$ with the property that no
vertex of $D$ is dominated by all the vertices of
$S_D$;\footnote{Note that this implies $d\geq 2$.}
\item if $D$ is an induced subgraph of an interval graph $G$ such that each
interval of $V(G)\setminus V(D)$ either contains all intervals of
$V(D)$ or does not intersect any of them, then for any solution $S$
for $G$, $|S\cap V(D)|\geq d$.\footnote{By this property, an interval of
$V(G)\setminus V(D)$ may only be useful to dominate a vertex in
$D$ (but not to separate a pair in $D$). Hence the property holds
if any optimal solution for the separation property only, has the same size as an optimal solution for both separation and domination.}
\end{itemize}
\end{defi}

\begin{figure}[!ht]
\centering
\begin{tikzpicture}[join=bevel,inner sep=0.5mm,line width=0.8pt, scale=0.4]

 \path (0,0.7) node (uv1) {};
 \path (2,-0.7) node (uv2) {};
\draw (-1,0) -- ++(12,0)
      (7.5,0.7) -- ++(3,0)
      (uv1) -- ++(7,0) node (uv1f) {}
      (uv2) -- ++(7,0) node (uv2f) {};
\path (4.5,-2.5) node (Puv) {$D$};
  
  \draw[rounded corners] 
                       (Puv)+(-1.3,1) rectangle ++(1.3,-1);
\end{tikzpicture}

\caption{Representation of dominating gadget $D$.}
\label{fig:D-gadget}
\end{figure}
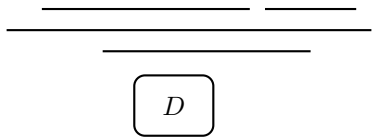

In the following, a dominating gadget will be represented graphically
as shown in Figure~\ref{fig:D-gadget}, where $D$ is an induced subgraph of an
interval graph $G$. In our constructions, we will build a graph $G$
with many isomorphic copies of $D$ as its induced subgraphs, where $D$
will be a fixed graph. Denote by $S$ an optimal solution for $G$: the
size of each local optimal solution $S\cap V(D)$ for $D$ will always
be~$d$. Moreover, the conditions of the second property in
Definition~\ref{def:domgadget} (that each interval of $V(G)\setminus
V(D)$ either contains all intervals of $V(D)$ or does not intersect
any of them) will always be satisfied.

\begin{claimbis}\label{claim:domgadget}
Let $G$ be an interval graph containing a dominating gadget $D$ as an
induced subgraph, such that each interval of $V(G)\setminus V(D)$
either contains all intervals of $V(D)$ or does not intersect any of
them. Then, for any optimal solution $S$ of $G$, $|S\cap V(D)|=d$ and we
can obtain an optimal solution $S'$ with $|S'|=|S|$ by replacing
$S\cap V(D)$ by the standard solution $S_D$.
\end{claimbis}
\begin{proof}
By the second property of a dominating gadget, we have $|S\cap
V(D)|\geq d\geq 1$. Since each interval of $V(G)\setminus V(D)$ either
contains all intervals of $V(D)$ or does not intersect any of them, a
pair of intervals of $V(G)\setminus V(D)$ either cannot be separated
by any interval in $V(D)$, or is separated equally by any interval in
$V(D)$. Since $d\geq 1$, there is at least one interval in $S\cap
V(D)$ but the structure of $S\cap D$ does not influence the rest of
the graph. Hence, $S\cap V(D)$ can be replaced by $S_D$ and we have $|S\cap V(D)|\leq d$ (otherwise the solution with $S_D$ would be better and $S$ would not be optimal).
\end{proof}

\begin{defi}[Choice pair]
A pair $\{u,v\}$ of intervals is called {\em choice pair} if $u,v$ both contain the intervals of a common dominating gadget (denoted $D(uv)$), and such that none of $u,v$ contains the other.
\end{defi}

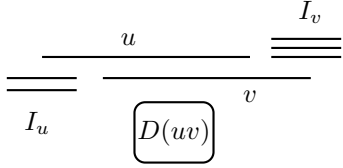
\begin{figure}[!ht]
\centering
\begin{tikzpicture}[join=bevel,inner sep=0.5mm,line width=0.8pt, scale=0.4]

 \path (0,0) node (uv1) {};
  \path (uv1)+(3,0.6) node {$u$};
  \path (uv1)+(2,-0.7) node (uv2) {};
  \path (uv2)+(5,-0.6) node {$v$};
  \path (uv2)+(5,2) node (v) {};
\draw (uv1) -- ++(7,0) node (uv1f) {}
      (uv2) -- ++(7,0) node (uv2f) {};
  \path (uv1)+(4.5,-2.5) node (Puv) {$D(uv)$};
  
  \draw[rounded corners] 
                       (Puv)+(-1.3,1) rectangle ++(1.3,-1);
                       
\draw (uv2)+(-0.7,0) -- ++(-3,0);
\draw (uv2)+(-0.7,-0.4) -- ++(-3,-0.4);
\path (uv1)+(0,-2.2) node {$I_u$};
\draw (uv1f)+(0.7,0) -- ++(3,0);
\draw (uv1f)+(0.7,0.3) -- ++(3,0.3);
\draw (uv1f)+(0.7,0.6) -- ++(3,0.6);
\path (uv2f)+(0,2.2) node {$I_v$};
\end{tikzpicture}
\caption{Choice pair $u,v$.}
\label{fig:choice}
\end{figure}

See Figure~\ref{fig:choice} for an illustration of a choice pair.
Intuitively, a choice pair gives us the choice of separating it from
the left or from the right: since none of $u,v$ is included in the
other, the intervals intersecting $u$ but not $v$ (the set $I_{u,v}$)
can only be located at one side of $u$; the same holds for $v$. In our
construction, we will make sure that all pairs of intervals will be
easily separated using domination gadgets. It will remain to separate
the choice pairs.

We have the following claim:

\begin{claimbis}\label{lem:choice}
Let $S$ be a solution of a graph $G$ and $\{u,v\}$ be a choice pair in
$G$. If the solution $S\cap V(D(uv))$ for the dominating gadget $D(uv)$ is
the standard solution $S_D$, both vertices $u$ and $v$ are
dominated, separated from all vertices in $D(uv)$ and from all
vertices not intersecting $D(uv)$.
\end{claimbis}
\begin{proof}
If $S$ is such a solution, by the definition of a dominating gadget,
$|S\cap D(uv)|\geq d\geq 1$. Since all vertices of $D(uv)$ are in the
open neighbourhood of $u$ and $v$, by
Property~\ref{prop:domsep}(1)-(3), $u$ and $v$ are dominated and
separated from the vertices not intersecting $D(uv)$. Moreover, both
$u,v$ are adjacent to all vertices of $D(uv)\cap S$. By
Definition~\ref{def:domgadget}, no vertex of $D(uv)$ is dominated by
the whole set $S_D$, hence $u,v$ are separated from all vertices in
$D(uv)$.
\end{proof}

We now define the central gadget of the reduction, the
\emph{transmitter gadget}. Roughly speaking, it allows to transmit
information across an interval graph using the separation property.

\begin{defi}[Transmitter gadget]\label{def:trans-gadget}
Let $P$ be a set of two or three choice pairs in an interval graph
$G$. A \emph{transmitter gadget} $Tr(P)$ is an induced subgraph of $G$ consisting of a path on seven vertices $\{u,uv^1,uv^2,v,vw^1,vw^2,w\}$ and five dominating gadgets $D(u)$, $D(uv)$, $D(v)$, $D(vw)$, $D(w)$ such that the following properties are satisfied:
\begin{itemize}
\item $u$ and $w$ are the only vertices of $Tr(P)$ that separate the pairs of $P$;
\item the intervals of the dominating gadget $D(u)$ (resp. $D(v)$, $D(w)$) are included in interval $u$ (resp. $v$, $w$) and no interval of $Tr(P)$ other than $u$ (resp. $v$, $w$) intersects $D(u)$ (resp. $D(v)$, $D(w)$);
\item pair $\{uv^1,uv^2\}$ is a choice pair and no interval of $V(Tr(P))\setminus (D(uv^1,uv^2)\cup \{uv^1,uv^2\})$ intersects both intervals of the pair. The same holds for pair $\{vw^1,vw^2\}$.
\item the choice pairs $\{uv^1,uv^2\}$ and $\{vw^1,vw^2\}$ cannot be separated by intervals of $G$ other than $u$, $v$ and $w$.
\end{itemize}
\end{defi}

\begin{figure}[!ht]
\centering
\scalebox{0.9}{\begin{tikzpicture}[join=bevel,inner sep=0.5mm,line width=0.8pt, scale=0.4]
  \path (-3,0.2) node (x1) {};
  \path (x1)+(1.25,0.5) node {$x_1$};
  \path (x1)+(2,1) node (x2) {};
  \path (x2)+(1.5,0.5) node {$x_2$};
  \path (1.1,-1) node (u) {};
  \path (u)+(3,0.5) node {$u$};
  \path (u)+(5.3,-1) node (uv1) {};
  \path (uv1)+(3,0.6) node {$uv^1$};
  \path (uv1)+(1.5,-1) node (uv2) {};
  \path (uv2)+(5,-0.6) node {$uv^2$};
  \path (uv2)+(5,2) node (v) {};
  \path (v)+(4,0.5) node {$v$};
  \path (v)+(11.5,1.2) node (y1) {};
  \path (y1)+(1.25,0.5) node {$y_1$};
  \path (y1)+(2,1) node (y2) {};
  \path (y2)+(1.5,0.5) node {$y_2$};
  \path (v)+(5,-1) node (vw1) {};
  \path (vw1)+(5,0.6) node {$vw^1$};
  \path (vw1)+(2.5,-1) node (vw2) {};
  \path (vw2)+(6.5,-0.6) node {$vw^2$};
  \path (vw2)+(8,2) node (w) {};
  \path (w)+(3,0.5) node {$w$};
  \path (w)+(5,1.2) node (z1) {};
  \path (z1)+(1.25,0.5) node {$z_1$};
  \path (z1)+(2,1) node (z2) {};
  \path (z2)+(1.5,0.5) node {$z_2$};
\path (u)+(3,-1.5) node (Pu) {$D(u)$};
\path (uv1)+(3.5,-2.5) node (Puv) {$D(uv)$};
\path (v)+(3,-1.5) node (Pv) {$D(v)$};
\path (vw1)+(5,-2.5) node (Pvw) {$D(vw)$};
\path (w)+(3.5,-1.5) node (Pw) {$D(w)$};
\draw (x1) -- ++(3,0)
      (x2) -- ++(3,0)
      (u) -- ++(6,0)
      (uv1) -- ++(6,0)
      (uv2) -- ++(6,0)
      (v) -- ++(7,0)
      (vw1) -- ++(10,0)
      (vw2) -- ++(9,0)
      (y1) -- ++(3,0)
      (y2) -- ++(3,0)
      (z1) -- ++(3,0)
      (z2) -- ++(3,0)
      (w) -- ++(6,0);
\draw[densely dashed] (x1)+(2.5,0.7) ellipse (3.2cm and 1.6cm);
\draw[densely dashed] (y1)+(2.5,0.7) ellipse (3.2cm and 1.6cm);
\draw[densely dashed] (z1)+(2.5,0.7) ellipse (3.2cm and 1.6cm);
\draw[rounded corners] (Pu)+(-1.2,1) rectangle ++(1.2,-1)
                       (Puv)+(-1.3,1) rectangle ++(1.3,-1)
                       (Pv)+(-1.2,1) rectangle ++(1.2,-1)
                       (Pvw)+(-1.3,1) rectangle ++(1.3,-1)
                       (Pw)+(-1.2,1) rectangle ++(1.2,-1);
   \path (v)+(4,-4) node (arrow) {};
   \draw [line width=2pt,->] (arrow) -- +(0,-1.5);
\path (17,-8.75) node {$Tr(\{x_1,x_2\},\{y_1,y_2\},\{z_1,z_2\})$};
\draw (0.5,-7) -- ++(1,0) -- ++(0,-1) -- ++(26.5,0) -- ++(0,1) --  ++(1,0) -- ++(0,-1) -- ++(5,0) -- ++(0,1) -- ++ (1,0) -- ++(0,-2.5) -- ++(-34.5,0) -- (0.5,-7);

\end{tikzpicture}}
\caption{Transmitter gadget $Tr(\{x_1,x_2\},\{y_1,y_2\},\{z_1,z_2\})$ and its ``box'' representation.
}
\label{fig:trans-gadget}
\end{figure}
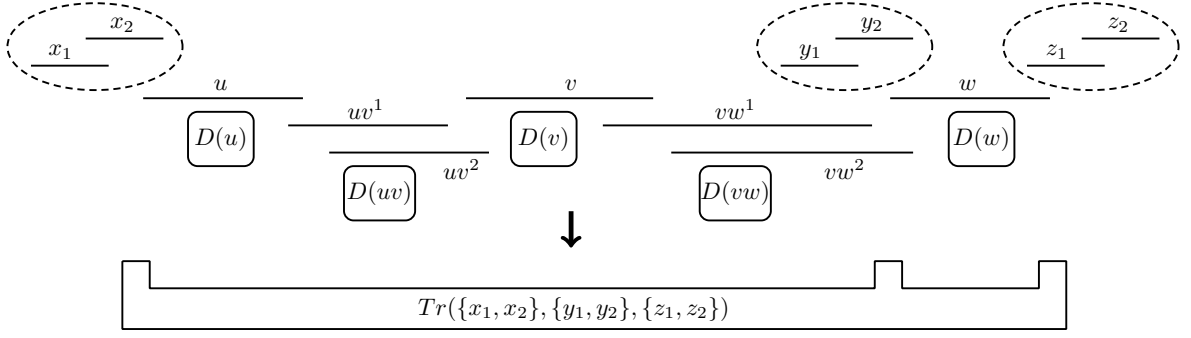

Figure~\ref{fig:trans-gadget} illustrates a transmitter gadget and shows
the succinct graphical representation that we will use. As shown in the figure, we may use a ``box'' to denote $T_r(P)$. This box does not include the choice pairs of $P$ but indicates where they are situated. Note that the middle pair $\{y_1,y_2\}$ could also be separated (from the left) by $u$ instead of $w$, or it may not exist at all.

The following claim shows how transmitter gadgets will be used in the
main reduction.

\begin{claimbis}\label{claim:trans}
Let $G$ be an interval graph with a transmitter gadget $Tr(P)$ and let
$S$ be a solution. We have $|S\cap Tr(P)|\geq 5d+1$ and if $|S\cap
Tr(P)|=5d+1$, no pair of $P$ is separated by a vertex in $S\cap
Tr(P)$.

Moreover, there exist two sets of vertices of $Tr(P)$, $S^-_{Tr(P)}$
and $S^+_{Tr(P)}$ of size $5d+1$ and $5d+2$ respectively, such that
the following holds.
\begin{itemize}
\item The set $S^-_{Tr(P)}$ dominates all the vertices of $Tr(P)$ and separates all the pairs of $Tr(P)$ but no pairs in $P$.
\item The set $S^+_{Tr(P)}$ dominates all the vertices of $Tr(P)$, separates all the pairs of $Tr(P)$ and all the pairs in $P$. 
\end{itemize}
\end{claimbis}
\begin{proof}
By the definition of the dominating gadget, we must have $|S\cap
Tr(P)|\geq 5d$ with $5d$ vertices of $S$ belonging to the dominating
gadgets. By Property~\ref{prop:domsep}(4) on the choice pair
$\{uv^1,uv^2\}$, at least one vertex of $\{u,uv^1,uv^2,v\}$ belongs to
$S$ (recall that the intervals not in $Tr(P)$ cannot separate the
choice pairs in $Tr(P)$), and similarly, for the choice pair
$\{vw^1,vw^2\}$, at least one vertex of $\{v, vw^1,vw^2,w\}$ belongs
to $S$. Hence $|S\cap Tr(P)|\geq 5d+1$ and if $|S\cap Tr(P)|=5d+1$,
vertex $v$ must be in $S$ and neither $u$ nor $w$ are in
$S$. Therefore, no pair of $P$ is separated by a vertex in $S\cap
Tr(P)$.

We now prove the second part of the claim. Let $S_{dom}$ be the union
of the five standard solutions $S_D$ of the dominating gadgets of
$Tr(P)$. Let $S^-_{Tr(P)}=S_{dom}\cup \{v\}$ and
$S^+_{Tr(P)}=S_{dom}\cup \{u,w\}$. The set $S_{dom}$ has $5d$ vertices
and so $S^-_{Tr(P)}$ and $S^+_{Tr(P)}$ have respectively $5d+1$ and
$5d+2$ vertices. Each interval of $Tr(P)$ either contains a
dominating gadget or is part of a dominating gadget and is therefore
dominated by a vertex in $S_{dom}$. Hence, pairs of vertices that are
not intersecting the same dominating gadget are clearly separated. By
the first property in Definition~\ref{def:domgadget}, a vertex
adjacent to a whole dominating gadget is separated from all the
vertices of the dominating gadget. Similarly, by definition, pairs of
vertices inside a dominating gadget are separated by
$S_{dom}$. Therefore, the only remaining pairs to consider are the
choice pairs. By Property~\ref{prop:domsep}(3), they are separated
both at the same time either by $v$ or by $\{u,w\}$. Hence the two sets
$S^-_{Tr(P)}$ and $S^+_{Tr(P)}$ are both dominating and separating the
vertices of $Tr(P)$. Moreover, since $S^+_{Tr(P)}$ contains $u$ and
$w$, it also separates the pairs of $P$.
\end{proof}

A transmitter gadget with a solution set of size $5d+1$ (resp. $5d+2$ vertices) is said to be {\em tight} (resp. {\em non-tight}). We
will call the sets $S^-_{Tr(P)}$ and $S^+_{Tr(P)}$ the {\em tight} and
\emph{non-tight standard solutions} of $Tr(P)$.

\subsection{The main reduction}\label{sec:reduc}

We are now ready to describe the reduction from \textsc{3-Dimensional Matching}. Each element $x\in A\cup B\cup C$ is modelled by a choice pair $\{f_x,g_x\}$. Each triple of $\mathcal T$ is modelled by a triple gadget defined as follows.

\begin{defi}[Triple gadget]\label{def:triple-gadget}
Let $T=\{a,b,c\}$ be a triple of $\mathcal
T$. The \emph{triple gadget} $G_t(T)$ is an interval graph consisting of four choice pairs $p=\{p_1,p_2\}$,
$q=\{q_1,q_2\}$, $r=\{r_1,r_2\}$, $s=\{s_1,s_2\}$ together with their associated dominating gadgets $D(p)$, $D(q)$, $D(r)$, $D(s)$ and five transmitter gadgets $Tr(p,q)$, $Tr(r,s)$, $Tr(s,a)$, $Tr(p,r,b)$ and $Tr(q,r,c)$, where:
\begin{itemize}
\item
$a=\{f_{a},g_{a}\}$, $b=\{f_{b},g_{b}\}$ and $c=\{f_{c},g_{c}\}$; 
\item Except for the choice pairs $p$, $q$, $r$, $s$, $a$, $b$, $c$, for each pair of intervals of $G_t(T)$, its two intervals intersect different subsets of dominating gadgets of $G_t(T)$.
\item In each transmitter gadget $Tr(P)$ and for each choice pair $\pi\in P$, the intervals of $\pi$ intersect the same intervals except for the vertices $u,v,w$ of $Tr(P)$;\item The intervals of $V(G)\setminus V(G_t(T))$ that are intersecting only a part of the gadget intersect accordingly to the transmitter gadget definition and do not separate the choice pairs $p$, $q$, $r$ and $s$.
\end{itemize}
\end{defi}

\begin{figure}[!ht]
\centering
\scalebox{0.74}{
\begin{tikzpicture}[join=bevel,inner sep=0.5mm,line width=0.8pt, scale=0.375]
  \path (0,0) node (p1) {};
  \path (p1)+(1,0.5) node {$p_1$};
  \path (p1)+(1.5,1) node (p2) {};
  \path (p2)+(1.5,0.5) node {$p_2$};

  \path (p1)+(10,0) node (q1) {};
  \path (q1)+(1,0.5) node {$q_1$};
  \path (q1)+(1.5,1) node (q2) {};
  \path (q2)+(1.5,0.5) node {$q_2$};

  \path (q1)+(6.5,0) node (r1) {};
  \path (r1)+(1,0.5) node {$r_1$};
  \path (r1)+(1.5,1) node (r2) {};
  \path (r2)+(1.5,0.5) node {$r_2$};

  \path (r1)+(10,0) node (s1) {};
  \path (s1)+(1,0.5) node {$s_1$};
  \path (s1)+(1.5,1) node (s2) {};
  \path (s2)+(1.5,0.5) node {$s_2$};

  \path (s1)+(10,0) node (a1) {};
  \path (a1)+(1,0.5) node {$f_{a}$};
  \path (a1)+(1.5,1) node (a2) {};
  \path (a2)+(1.5,0.5) node {$g_{a}$};

  \path (a1)+(7,0) node (b1) {};
  \path (b1)+(1,0.5) node {$f_{b}$};
  \path (b1)+(1.5,1) node (b2) {};
  \path (b2)+(1.5,0.5) node {$g_{b}$};

  \path (b1)+(7,0) node (c1) {};
  \path (c1)+(1,0.5) node {$f_{c}$};
  \path (c1)+(1.5,1) node (c2) {};
  \path (c2)+(1.5,0.5) node {$g_{c}$};

\path (s1)+(8,0) node {\ldots};

\path (p1)+(2.8,-1.5) node (Pp) {\small{$D(p)$}};
\path (q1)+(2.8,-1.5) node (Pq) {\small{$D(q)$}};
\path (r1)+(2.8,-1.5) node (Pr) {\small{$D(r)$}};
\path (s1)+(2.8,-1.5) node (Ps) {\small{$D(s)$}};
\path (a1)+(3.2,-1.5) node (Pa) {\small{$D(a)$}};
\path (b1)+(3.2,-1.5) node (Pb) {\small{$D(b)$}};
\path (c1)+(3.2,-1.5) node (Pc) {\small{$D(c)$}};

\draw (p1) -- ++(4,0)
      (p2) -- ++(4,0)
      (q1) -- ++(4,0)
      (q2) -- ++(4,0)
      (r1) -- ++(4,0)
      (r2) -- ++(4,0)
      (s1) -- ++(4,0)
      (s2) -- ++(4,0)
      (a1) -- ++(4.5,0)
      (a2) -- ++(4.5,0)
      (b1) -- ++(4.5,0)
      (b2) -- ++(4.5,0)
      (c1) -- ++(4.5,0)
      (c2) -- ++(4.5,0);

\draw[rounded corners] (Pp)+(-1.0,0.8) rectangle ++(1.0,-0.8)
                       (Pq)+(-1.0,0.8) rectangle ++(1.0,-0.8)
                       (Pr)+(-1.0,0.8) rectangle ++(1.0,-0.8)
                       (Ps)+(-1.0,0.8) rectangle ++(1.0,-0.8)
                       (Pa)+(-1.0,0.8) rectangle ++(1.0,-0.8)
                       (Pb)+(-1.0,0.8) rectangle ++(1.0,-0.8)
                       (Pc)+(-1.0,0.8) rectangle ++(1.0,-0.8);

\path (8,-2.65) node {$Tr(p,q)$};
\path (24.5,-2.65) node {$Tr(r,s)$};
\path (27,3.75) node {$Tr(p,r,b)$};
\path (33,6.95) node {$Tr(q,r,c)$};

\path (34.5,10.15) node {$Tr(s,a)$};

\draw (4.5,-0.8) -- ++(1,0) -- ++(0,-1) -- ++(3,0) -- ++(2,0) -- ++(0,1) -- ++ (1,0) -- ++(0,-2.5) -- ++(-7,0) -- (4.5,-0.8);

\draw (21,-0.8) -- ++(1,0) -- ++(0,-1) -- ++(3,0)  -- ++(2,0) -- ++(0,1) -- ++ (1,0) -- ++(0,-2.5) -- ++(-7,0) -- (21,-0.8);

\draw (4.5,2) -- ++(1,0) -- ++(0,1) -- ++(11,0) -- ++(0,-1) --  ++(1,0) -- ++(0,1) -- ++(26,0) -- ++(0,-1) -- ++ (1,0) -- ++(0,2.5) -- ++(-40,0) -- cycle;

\draw (14.5,5.2) -- ++(1,0) -- ++(0,1) -- ++(5,0) -- ++(0,-1) --  ++(1,0) -- ++(0,1) -- ++(29.25,0) -- ++(0,-1) -- ++ (1,0) -- ++(0,2.5) -- ++(-37.25,0) -- cycle;

\draw (31.25,8.4) -- ++(1,0) -- ++(0,1) -- ++(2,0) -- ++(2,0) -- ++(0,-1) -- ++ (1,0) -- ++(0,2.5) -- ++(-6,0) --cycle;

\end{tikzpicture}
}
\caption{Triple gadget $G_t(\{a,b,c\})$ together with choice pairs of elements $a$, $b$ and $c$.}
\label{fig:triple-gadget}
\end{figure}
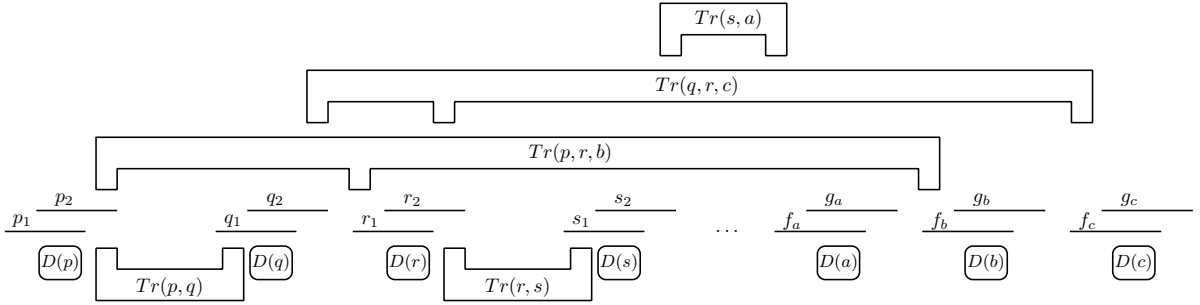

Note that there are several ways to obtain a triple gadget that is an interval graph and that satisfies the properties in Definition~\ref{def:triple-gadget}. The one in Figure~\ref{fig:triple-gadget} represents one of these possibilities. We remark that $p$, $q$, $r$ and $s$
in $G_t(\{a,b,c\})$, are all functions of $\{a,b,c\}$ but to simplify
the notations we simply write $p$, $q$, $r$ and $s$.

\begin{claimbis}\label{claim:triple}
Let $G$ be a graph with a triple gadget $G_t(T)$ and $S$ be a
solution. We have $|S\cap G_t(T)|\geq 29d + 7$ and if $|S\cap
G_t(T)|=29d+7$, no choice pair corresponding to $a$, $b$ or $c$ is
separated by a vertex in $S\cap G_t(T)$.

Moreover, there exist two sets of vertices of $G_t(T)$, $S^-_{G_t(T)}$ and $S^+_{G_t(T)}$ of size $29d+7$ and $29d+8$ respectively, such that the following holds.
\begin{itemize}
\item The set $S^-_{G_t(T)}$ dominates all the vertices of $G_t(T)$ and separates all the pairs of $G_t(T)$ but does not separate any choice pairs corresponding to $\{a,b,c\}$.
\item The set $S^+_{G_t(T)}$ dominates all the vertices of $G_t(T)$, separates all the pairs of $G_t(T)$ and separates the choice pairs corresponding to $\{a,b,c\}$.
\end{itemize}
\end{claimbis}
\begin{proof}
The proof is similar of the proof of Claim~\ref{claim:trans}. Each
transmitter gadget must contain at least $5d+1$ vertices, and each of
the four dominating gadgets of the choice pairs $p$, $q$, $r$, $s$ must
contain $d$ vertices. Hence there must be already $29d + 5$ vertices
of $G_t(T)$ in the solution. Furthermore, to separate the choice pair
$s$, $Tr(r,s)$ or $Tr(s,a)$ must be non-tight (since $s$
is not separated by other vertices of the graph). In the same way,
to separate the choice pair $p$, $Tr(p,q)$ or $Tr(p,r,b)$ must be
non-tight. Then at least two transmitter gadgets are non-tight and we
have $|S\cap G_t(T)|\geq 29d + 7$. If $|S\cap G_t(T)|=29d + 7$,
exactly two transmitter gadgets are non-tight and they can only be
$Tr(r,s)$ and $Tr(p,q)$ (otherwise some of the choice pairs $p,q,r,s$ would not be separated). Hence the choice pairs corresponding to
$\{a,b,c\}$ are not separated by the vertices of $G_t(T)\cap S$.

For the second part of the claim, the set $S^-_{G_t(T)}$ is defined by
taking the union of the tight standard solutions of $Tr(s,a)$,
$Tr(q,r,c)$ and $Tr(p,r,b))$, the non-tight standard solutions of
$Tr(p,q)$ and $Tr(r,s)$ and the standard solutions of the dominating
gadgets $D(p)$, $D(q)$, $D(r)$ and $D(s)$. The set $S^+_{G_t(T)}$ is
defined by taking the union of the non-tight standard solutions of
$Tr(s,a)$, $Tr(q,r,c)$ and $Tr(p,r,b)$, the tight standard solutions
of $Tr(p,q)$ and $Tr(r,s)$ and the standard solutions of the
dominating gadgets $D(p)$, $D(q)$, $D(r)$ and $D(s)$. By Claim~\ref{claim:trans}, the definition of a dominating gadget and the fact
that the only intervals sharing the same sets of dominating gadgets
are the choice pairs, all intervals of $G_t(T)$ are dominated and all
the pairs of intervals except the choice pairs are separated by both
$S^-_{G_t(T)}$ and $S^+_{G_t(T)}$. The choice pairs $p$, $q$, $r$ and
$s$ are separated by the non-tight solutions of the transmitter
gadgets. Hence $S^-_{G_t(T)}$ and $S^+_{G_t(T)}$ are dominating and
separating all the intervals of $G_t(T)$.

When the solution contains $S^-_{G_t(T)}$, the transmitter gadgets
$Tr(s,a)$, $Tr(q,r,c)$ and $Tr(p,r,b)$ are tight. Hence $S^-_{G_t(T)}$
does not separate any choice pairs among $\{a,b,c\}$. On the other
hand, since $S^+_{G_t(T)}$ contains the non-tight solution of
$Tr(s,a)$, $Tr(q,r,c)$ and $Tr(p,r,b)$, the three choice pairs
$\{a,b,c\}$ are separated by $S^+_{G_t(T)}$.
\end{proof}

As before, a triple gadget with $29d+7$ vertices (resp. $29d+8$) is said to be {\em tight} (resp. {\em non-tight}). We will call the sets $S^-_{G_t(T)}$ and $S^+_{G_t(T)}$ the {\em tight} and {\em non-tight standard solutions} of $G_t(T)$.

\begin{figure}[!h]
\includegraphics[scale=0.85]{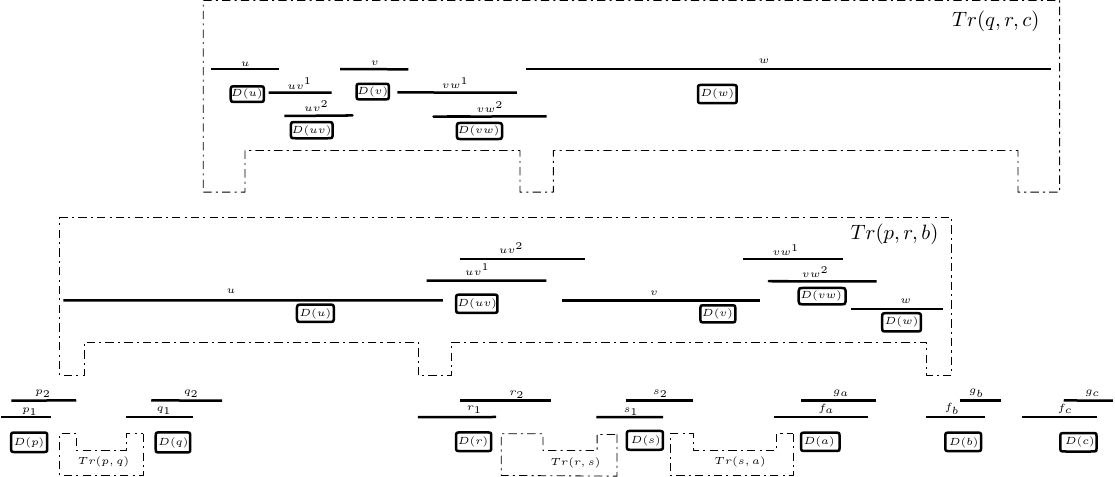}
\caption{The detailed construction of a triple gadget.}
\label{fig:triple-gadget-detailed}
\end{figure}

Given an instance $(A, B, C, \mathcal{T})$ of \textsc{3-Dimensional Matching} with $|A|=|B|=|C|=n$ and $|\mathcal{T}|=m$, we construct the interval graph $G=G(A, B, C,\mathcal{T})$ as follows.
\begin{itemize}
\item As mentioned previously, to each element $x$ of $A\cup B\cup C$, we assign a distinct choice pair $\{f_x,g_x\}$ in $G$. The intervals of any two distinct choice pairs $\{f_x,g_x\},\{f_y,g_y\}$ are disjoint and they are all in $\mathbb{R}^+$.

\item For each triple $T=\{a,b,c\}$ of $\mathcal T$ we first associate an interval $I_T$ in $\mathbb{R}^-$ such that for any two triples $T_1$ and $T_2$, $I_{T_1}$ and $I_{T_2}$ do not intersect. Then inside $I_T$, we build the choice pairs $\{p_1,p_2\}$, $\{q_1,q_2\}$, $\{r_1,r_2\}$, $\{s_1,s_2\}$. Finally, using the choice pairs already associated to elements $a$, $b$ and $c$ we complete this to a triple gadget.

\item When placing the remaining intervals of the triple gadgets, we must ensure that triple gadgets do not ``interfere'': for every dominating gadget $D$, no interval in $V(G)\setminus V(D)$ must have an endpoint inside $D$. Similarly, the choice pairs of every triple gadget or transmitter gadget must only be separated by intervals among $u$, $v$ and $w$ of its corresponding private transmitter gadget.

For intervals of distinct triple gadgets, this is easily done by our placement of the triple gadgets. To ensure that the intervals of transmitter gadgets of the same triple gadget do not interfere, we proceed as follows. We place the whole gadget $Tr(p,q)$ inside interval $u$ of $Tr(p,r,b)$. Similarly, the whole $Tr(r,s)$ is placed inside interval $w$ of $Tr(q,r,c)$ and the whole $Tr(s,a)$ is placed inside interval $v$ of $Tr(p,r,b)$. One has to be more careful when placing the intervals of $Tr(p,r,b)$ and $Tr(q,r,c)$. In $Tr(p,r,b)$, we must have that interval $u$ separates $p$ from the right of $p$. We also place $u$ so that it separates $r$ from the left of $r$. Intervals $uv^1,uv^2$ both start in $r_1$, so that $u$ also separates $uv^1,uv^2$ and also to ensure that $uv^1,uv^2$ does not separate the choice pair $r$.
Intervals $uv^1,uv^2$ continue until after pair $s$. In $Tr(q,r,c)$, we place $u$ so that it separates $q$ from the right, and we place $w$ so that it separates $r$ from the right; intervals $uv^1,uv^2,v$ lie strictly between $q$ and $r$; intervals $vw^1,vw^2$ intersect $r_1,r_2$ but stop before the end of $r_2$ (so that $w$ can separate both pairs $vw^1,vw^2$ and $r$ but without these pairs interfering). It is now easy to place $Tr(s,a)$ between $s$ and $a$. An example is given in Figure~\ref{fig:triple-gadget-detailed}.
\end{itemize}

The graph $G(A,B,C,\mathcal{T})$ has $(29v_D+43)m+3(v_D+2)n$ vertices (where $v_D$ is the order of a dominating gadget) and the interval representation described by our procedure can be obtained in polynomial time. We are now ready to state the main result of this section.

\begin{thm}\label{th:reduction}
$(A, B, C, \mathcal{T})$ has a perfect 3-dimensional matching if and only if $G=G(A,B,C, \mathcal{T})$ has a solution with $(29d+7)m+(3d+1)n$ vertices.
\end{thm}
\begin{proof}
Let $\mathcal M$ be a perfect 3-dimensional matching of $(A, B,
C, \mathcal{T})$. Let $S^+$ (resp. $S^-$) be the union of all the
non-tight (resp. tight) standard solutions $S^+_{G_t(T)}$ for $T\in
\mathcal M$ (resp. $S^-_{G_t(T)}$ for $T\notin \mathcal M$). Let $S_d$
be the union of all the standard solutions of the dominating gadgets
corresponding to the choice pairs of the elements.

Then $S=S^+\cup S^- \cup S_d$ is a solution of size
$(29d+7)m+(3d+1)n$. Indeed, by the definition of the dominating
gadgets, all the intervals inside a dominating gadget are dominated
and separated from all the other intervals. All the other intervals
intersect at least one dominating gadget and thus are
dominated. Furthermore, two intervals that are not a choice pair do
not intersect the same set of dominating gadgets and thus are
separated by one of the dominating gadgets. Finally, the choice pairs
inside a triple gadget are separated by Claim~\ref{claim:triple} and
the choice pairs corresponding to the elements of $A\cup B\cup C$ are
separated by the non-tight standard solutions of the triple gadgets
corresponding to the perfect matching.

\vspace{6mm}

Now, let $S$ be a solution of size $(29d+7)m+(3d+1)n$. We may assume that
the solution is standard on all triple gadgets and on the dominating
gadgets. Let $n_2$ be the number of non-tight triple gadgets in the
solution $S$. By Claim~\ref{claim:triple}, there must by at least
$(29d+7)m+n_2$ vertices of $S$ inside the $m$ triple gadgets and $3dn$
vertices of $S$ for the dominating gadgets of the $3n$ elements of
$A\cup B\cup C$. Hence $(29d+7)m+n_2+3dn\leq (29d+7)m+(3d+1)n$ and we
have $n_2\leq n$. Each non-tight triple gadget can separate three
choice pairs corresponding to the elements of $A\cup B\cup C$. Hence,
if $n_2<n$, it means that at least $3(n-n_2)$ choice pairs
corresponding to elements are not separated by a triple gadget. By the
separation property, the only way to separate a choice pair
$\{f_x,g_x\}$ without using a non-tight triple gadget is to have $f_x$
or $g_x$ in the solution. Hence we need $3(n-n_2)$ vertices to
separate these $3(n-n_2)$ choice pairs, and these vertices are not in
the triple gadgets nor in the dominating gadgets. Hence the solution
has size at least $(29d+7)m+n_2+3dn+3(n-n_2)>(29d+7)m+(3d+1)n$, leading to
a contradiction.

Therefore, $n_2=n$ and there are exactly $n$ non-tight triple gadgets.
Each of them separates three choice element pairs and since there are
$3n$ elements, the non-tight triple gadgets separate distinct choice
pairs. Hence, the set of triples $\mathcal M$ corresponding to the
non-tight triple gadgets is a perfect 3-dimensional matching of
$(A, B, C, \mathcal T)$.
\end{proof}

\begin{cor}
Any graph distinguishing problem $P$ based on domination and separation satisfying Property~\ref{prop:domsep} and admitting a dominating gadget that is an interval graph, is NP-complete even for the class of interval graphs.
\end{cor}

\medskip

A similar hardness result can be derived for the class of permutation graphs as follows.


\begin{cor}
Any graph distinguishing problem $P$ based on domination and
separation satisfying Property~\ref{prop:domsep} and admitting a
dominating gadget that is a permutation graph, is
NP-complete even for the class of permutation graphs.
\end{cor}

\begin{figure}[!ht]
\centering
\begin{tikzpicture}[join=bevel,inner sep=0.5mm,line width=0.8pt, scale=0.4]

\draw[help lines] (-5,2) -- (5,2) 
      (-5,-2) -- (5,-2);

\draw (-3,2) -- (3,-2);
\draw (-3,-2) -- (3,2);

\draw[rounded corners, fill=white] (0,0)+(-1,1.5) rectangle node[fill=white] {$D$} ++(1,-1.5);
\end{tikzpicture}
\caption{Representation of a dominating gadget as a permutation diagram.}
\label{fig:D-gadgetperm}
\end{figure}
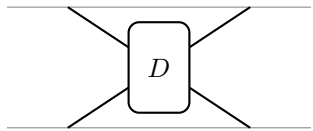

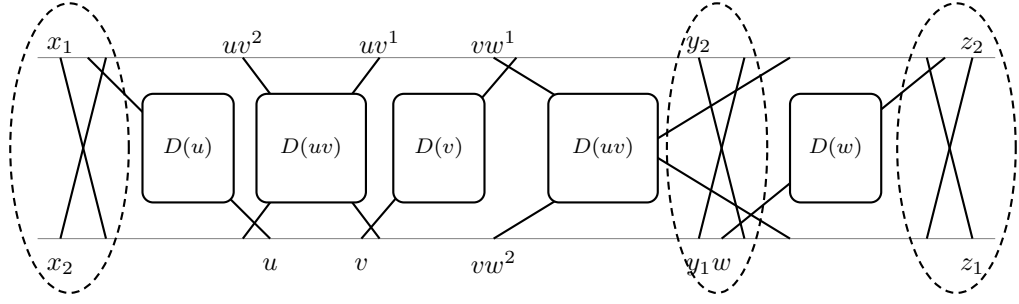
\begin{figure}[!ht]
\centering
\begin{tikzpicture}[join=bevel,inner sep=0.5mm,line width=0.8pt, scale=0.6]

\draw[help lines] (-10.5,2) -- (10.5,2) 
      (-10.5,-2) -- (10.5,-2);

\draw (-10,2) node[above] {$x_1$} -- (-9,-2);
\draw (-10,-2) node[below=0.2cm] {$x_2$} -- (-9,2) ;

\draw[densely dashed] (-9.8,0) ellipse (1.3cm and 3.2 cm);


\draw (-9.4,2)  -- (-5.4,-2) node[below=0.2cm] {$u$};
\draw[rounded corners, fill=white] (-7.2,0)+(-1,1.2) rectangle node[fill=white] {\footnotesize $D(u)$} ++(1,-1.2);

\draw (-6,-2) -- (-3,2)  node[above] {$uv^1$}  ;
\draw (-3,-2) -- (-6,2)  node[above] {$uv^2$}  ;
\draw[rounded corners, fill=white] (-4.5,0)+(-1.2,1.2) rectangle node[fill=white] {\footnotesize $D(uv)$} ++(1.2,-1.2);

\draw (-3.4,-2) node[below=0.2cm] {$v$} -- (0,2) ;
\draw[rounded corners, fill=white] (-1.7,0)+(-1,1.2) rectangle node[fill=white] {\footnotesize $D(v)$} ++(1,-1.2);

\draw (-0.5,-2) node[below=0.1cm] {$vw^2$}  -- (6,2)   ;
\draw (6,-2) -- (-0.5,2)  node[above] {$vw^1$}  ;
\draw[rounded corners, fill=white] (1.9,0)+(-1.2,1.2) rectangle node[fill=white] {\footnotesize $D(uv)$} ++(1.2,-1.2);

\draw (4.5,-2) node[below=0.2cm] {$w$} -- (9.4,2) ;
\draw[rounded corners, fill=white] (7,0)+(-1,1.2) rectangle node[fill=white] {\footnotesize $D(w)$} ++(1,-1.2);

\draw (4,2) node[above] {$y_2$} -- (5,-2) ;
\draw (4,-2)node[below=0.2cm] {$y_1$}  -- (5,2)  ;
\draw[densely dashed] (4.4,0) ellipse (1.1cm and 3.2 cm);

\draw (9,2) -- (10,-2) node[below=0.2cm] {$z_1$};
\draw (9,-2)  -- (10,2) node[above] {$z_2$} ;

\draw[densely dashed] (9.65,0) ellipse (1.3cm and 3.2 cm);

\end{tikzpicture}
\caption{Representation of a the transmitter gadget as a permutation diagram.}
\label{fig:transperm}
\end{figure}

\begin{figure}[!ht]
\centering

\tikzset{snake it/.style={decorate, decoration=snake},
    Trprbedge/.style = {color=gray},
    Trqrcedge/.style = {dashed,color=black},
    choicepairedge/.style = {decoration = {zigzag,segment length = 1.5mm, amplitude = .25mm},decorate,line width=0.5mm}
}

\begin{tikzpicture}[join=bevel,inner sep=0.5mm,line width=0.8pt, scale=0.12]
\node at (18,52) {};
\node at (18,46) { Dashed : $Tr(q,r,c)$};
\node at (18,41) { Gray : $Tr(p,r,b)$};

\draw[help lines] (-1.5,20) -- (132,20) 
      (-1.5,0) -- (132,0);

\draw[choicepairedge] (0,20)-- (4,0);
\draw[choicepairedge] (0,0)-- (9,20);
\node at (2,-3.5) {\scriptsize \textbf{pair $\bm p$}};

\draw[choicepairedge] (20,20) -- (24,0);
\draw[choicepairedge] (15,0) -- (25,20);
\node at (19,-3.5) {\scriptsize \textbf{pair $\bm q$}};

\draw[choicepairedge] (42,20) -- (63,0);
\draw[choicepairedge] (50,0) -- (62,20);
\node at (56,-3.5) {\scriptsize \textbf{pair $\bm r$}};

\draw[choicepairedge] (75,20) -- (88,0);
\draw[choicepairedge] (75,0) -- (88,20);
\node at (81.5,-3.5) {\scriptsize \textbf{pair $\bm s$}};

\draw[choicepairedge] (102,20) -- (109,0);
\draw[choicepairedge] (102,0) -- (109,20);
\node at (105.5,-3.5) {\scriptsize \textbf{pair $\bm a$}};

\draw[choicepairedge] (116,20) -- (122,0);
\draw[choicepairedge] (116,0) -- (122,20);
\node at (117,-3.5) {\scriptsize \textbf{pair $\bm b$}};

\draw[choicepairedge] (123,20) -- (130,0);
\draw[choicepairedge] (123,0) -- (130,20);
\node at (128,-3.5) {\scriptsize \textbf{pair $\bm c$}};

\draw[Trqrcedge] (27,20) node[above=1] (uv1Trqrc) {\scriptsize $uv^1$} -- (35,0);
\draw[dotted,color=black] (27,20) -- (uv1Trqrc);

\draw[Trqrcedge] (21,0) -- (31,20) node[above=1] (uTrqrc) {\scriptsize $u$} ;
\draw[dotted,color=black] (31,20) -- (uTrqrc);

\draw[Trqrcedge] (33.5,20) node[above=1] (vTrqrc) {\scriptsize $v$} -- (41.5,0);
\draw[dotted,color=black] (33.5,20) -- (vTrqrc);

\draw[Trqrcedge] (28,0) -- (37,20) node[above=1] (uv2Trqrc) {\scriptsize $uv^2$} ;
\draw[dotted,color=black] (37,20) -- (uv2Trqrc);

\draw[Trqrcedge] (54,20) node[above=1] (vw2Trqrc) {\scriptsize $vw^2$~~~} -- (57.5,0);
\draw[dotted,color=black] (54,20) -- (vw2Trqrc);

\draw[Trqrcedge] (38,0) -- (67,20) node[above=1] (vw1Trqrc) {\scriptsize $vw^1$} ;
\draw[dotted,color=black] (67,20) -- (vw1Trqrc);

\draw[Trqrcedge] (53,0) -- (127.5,20) node[above=1] (wTrqrc) {\scriptsize $w$} ;
\draw[dotted,color=black] (127.5,20) -- (wTrqrc);

\draw[Trprbedge] (59.5,0) -- (3,20) node[above=2] (uTrprb) {\scriptsize $u$} ;
\draw[dotted,color=gray] (3,20) -- (uTrprb);

\draw[Trprbedge] (69,0) -- (39,20) node[above=2] (uv2Trprb) {\scriptsize $uv^2$} ;
\draw[dotted,color=gray] (39,20) -- (uv2Trprb);

\draw[Trprbedge] (56,0) -- (56,20) node[above=2] (uv1Trprb) {\scriptsize $uv^1$} ;
\draw[dotted,color=gray] (56,20) -- (uv1Trprb);

\draw[Trprbedge] (61,0) -- (95.5,20) node[above=2] (vTrprb) {\scriptsize $v$} ;
\draw[dotted,color=gray] (95.5,20) -- (vTrprb);

\draw[Trprbedge] (113,0) -- (90.5,20) node[above=2] (vw1Trprb) {\scriptsize $vw^1$} ;
\draw[dotted,color=gray] (90.5,20) -- (vw1Trprb);

\draw[Trprbedge] (119,0) -- (111,20) node[above=2] (wTrprb) {\scriptsize $w$~~} ;
\draw[dotted,color=gray] (111,20) -- (wTrprb);

\draw[Trprbedge] (97,0) -- (114,20) node[above=2] (vw2Trprb) {~~~\scriptsize $vw^2$} ;
\draw[dotted,color=gray] (114,20) -- (vw2Trprb);

\draw[densely dotted] (5,4) -- (5,16) -- (19,16) -- (19,4) -- cycle;
\node at (12,9.5) {\small $Tr(p,q)$};

\draw[densely dotted] (53,12) -- (53,20) -- (79.5,20) -- (79.5,12) -- (73.5,12) -- (73.5,0) -- (65,0) -- (65,12) -- cycle;
\node at (69,14) {\small $Tr(r,s)$};

\draw[densely dotted] (85,12) -- (85,20) -- (105,20) -- (105,12) -- (100,12) -- (100,0) -- (90,0) -- (90,12) -- cycle;
\node at (95,6) {\small $Tr(s,a)$};
\end{tikzpicture}

\caption{Representation of a the triple gadget as a permutation diagram.}
\label{fig:tripleperm}
\end{figure}
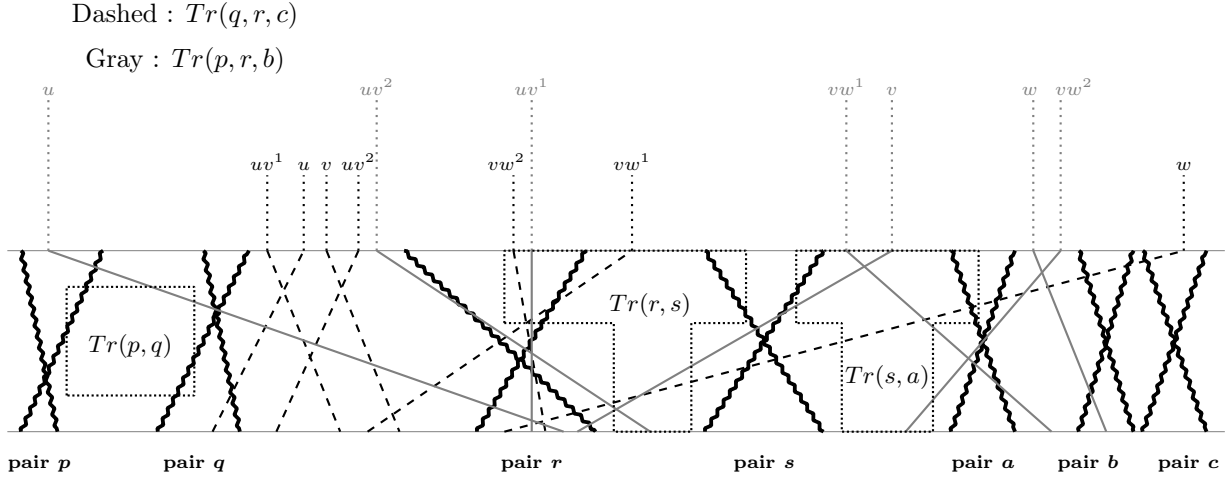

\begin{proof} We can use the same reduction as the one that yields Theorem~\ref{th:reduction}. We represent a permutation graph using its intersection model of segments as defined in the introduction. A dominating gadget will be represented as in Figure~\ref{fig:D-gadgetperm}. The transmitter gadget of Definition~\ref{def:trans-gadget} is also a permutation graph, see Figure~\ref{fig:transperm} for an illustration. Using these gadgets, we can build a triple gadget that satisfies Definition~\ref{def:triple-gadget} and is a permutation graph. A simplified permutation diagram (without dominating gadgets) of such a triple gadget is given in Figure~\ref{fig:tripleperm}.
Now, similarly as for interval graphs, given an instance $(A,B,C,\mathcal{T})$ of \textsc{3-Dimensional Matching}, one can define a graph $G=G(A,B,C, \mathcal{T})$ that is a permutation graph and for which $(A,B,C, \mathcal{T})$ has a perfect 3-dimensional matching if and only if $G=G(A,B,C, \mathcal{T})$ has a solution with $(29d + 7)m + (3d + 1)n$ vertices. The proof is the same as the one in Theorem~\ref{th:reduction}.
\end{proof}

\subsection{Applications to the specific problems}\label{sec:app}

We now apply Theorem~\ref{th:reduction} to \PBLD, \PBID{} and \PBOLD{}
by providing corresponding dominating gadgets.

\begin{cor}\label{cor:NPID}\label{coro:perm-inter-NPc}
\PBLD, \PBID{} and \PBOLD{} are NP-complete for interval graphs and permutation graphs.
\end{cor}
\begin{proof}
We prove that the path graphs $P_4$, $P_5$ and $P_6$ are dominating gadgets for \PBLD, \PBID{} and \PBOLD, respectively. These graphs are clearly interval and permutation graphs at the same time. To comply with Definition~\ref{def:domgadget}, we must prove that a dominating gadget $D$ (i) has an optimal solution $S_D$ of size~$d$ such that no vertex of $D$ is dominated by all the vertices of $S_D$, and (ii) if $D$ is an induced subgraph of an interval graph $G$ such that each interval of $V(G)\setminus V(D)$ either contains all intervals of $V(D)$ or does not intersect any of them, then for any solution $S$ for $G$, $|S\cap V(D)|\geq d$.

\begin{itemize}
\item \PBLD. Let $V(P_4)=\{x_1,\ldots, x_4\}$ and $d=2$. The set $S_D=\{x_1,x_4\}$ satisfies (i). For (ii), assume that $S$ is a locating-dominating set of a graph $G$ containing a copy $P$ of $P_4$ satisfying the conditions. If $S\cap P=\emptyset$ or $S\cap P=\{x_1\}$, then $x_3$ and $x_4$ are not separated. If $S\cap P=\{x_2\}$, then $x_1$ and $x_3$ are not separated. Hence, by symmetry, there at least two vertices of $P$ in $S$, and (ii) is satisfied.

\item\PBID. Let $V(P_5)=\{x_1,\ldots,x_5\}$ and $d=3$. The set $S_D=\{x_1,x_3,x_5\}$ satisfies (i). For (ii), assume that $S$ is an identifying code of a graph $G$ containing a copy $P$ of $P_5$ satisfying the conditions. To separate the pair $\{x_1,x_2\}$, we must have $x_3\in S$, since the other vertices cannot separate any pair inside $P$. To separate the pair $\{x_2,x_3\}$, we must have $\{x_1,x_4\}\cap S \neq \emptyset$ and to separate the pair $\{x_3,x_4\}$, we must have $\{x_2,x_5\}\cap S \neq \emptyset$. Hence there at least three vertices of $P$ in $S$, and (ii) is satisfied.

\item\PBOLD. Let $V(P_6)=\{x_1,\ldots, x_6\}$ and $d=4$. The set $S_D=\{x_1,x_3,x_4,x_6\}$ satisfies (i). For (ii), assume that $S$ is an open locating-dominating set of a graph $G$ containing a copy $P$ of $P_6$ satisfying the conditions. To separate the pair $\{x_1,x_3\}$, we
must have $x_4\in S$. Symmetrically, $x_3 \in S$. To separate the pair
$\{x_2,x_4\}$, we might have $\{x_1,x_5\}\cap S \neq \emptyset$ and
symmetrically, $\{x_2,x_6\}\cap S \neq \emptyset$. Hence there at
least four vertices of $P$ in $S$, and (ii) is satisfied.
\end{itemize}
\end{proof}

\subsection{Reductions for diameter~$2$ and consequence for \PBMD}\label{sec:diam2}

We now describe self-reductions for \PBID, \PBLD{} and \PBOLD{} for
graphs with a universal vertex (hence, graphs of diameter~$2$). We also
give a similar reduction from \PBLD{} to \PBMD.

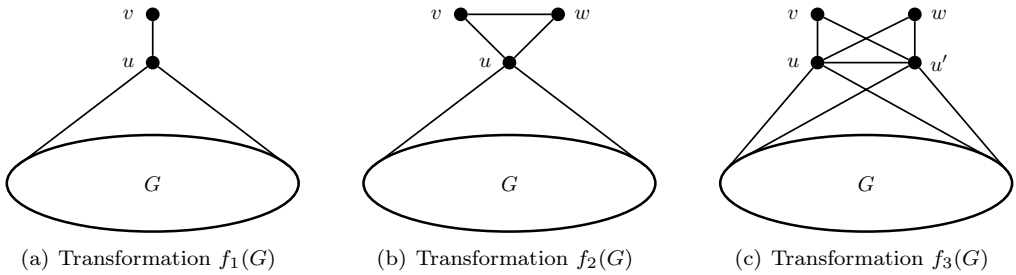
\begin{figure}[ht!]
\begin{center}
\subfigure[Transformation $f_1(G)$]{\label{fig:red-diam2-f1}
\scalebox{0.8}{\begin{tikzpicture}[join=bevel,inner sep=0.7mm,line width=0.8pt,scale=0.8]
\path (0,0) node (g) {};
\path (g)+(0,-0.5) node {$G$};
  \path (g)+(0,2) node[draw,shape=circle,fill] (u) {};
  \path (u)+(-0.5,0) node {$u$};
  \path (u)+(0,1) node[draw,shape=circle,fill] (v) {};  
  \path (v)+(-0.5,0) node {$v$};

  \draw (u) -- (v)
(g)+(-2.8,-0.15) -- (u)
(g)+(2.8,-0.15) -- (u);

  \draw[line width=1.2pt] (g)+(0,-0.5) ellipse (3cm and 1cm);

\end{tikzpicture}}}\qquad
\subfigure[Transformation $f_2(G)$]{\label{fig:red-diam2-f2}
\scalebox{0.8}{\begin{tikzpicture}[join=bevel,inner sep=0.7mm,line width=0.8pt,scale=0.8]
\path (0,0) node (g) {};
\path (g)+(0,-0.5) node {$G$};
  \path (g)+(0,2) node[draw,shape=circle,fill] (u) {};
  \path (u)+(-0.5,0) node {$u$};
  \path (u)+(-1,1) node[draw,shape=circle,fill] (v) {};  
  \path (v)+(-0.5,0) node {$v$};
  \path (u)+(1,1) node[draw,shape=circle,fill] (w) {};  
  \path (w)+(0.5,0) node {$w$};
  \draw (u) -- (v) -- (w) -- (u)
(g)+(-2.8,-0.15) -- (u)
(g)+(2.8,-0.15) -- (u);

  \draw[line width=1.2pt] (g)+(0,-0.5) ellipse (3cm and 1cm);

\end{tikzpicture}}}\qquad
\subfigure[Transformation $f_3(G)$]{\label{fig:red-diam2-f3}
\scalebox{0.8}{\begin{tikzpicture}[join=bevel,inner sep=0.7mm,line width=0.8pt,scale=0.8]
\path (0,0) node (g) {};
\path (g)+(0,-0.5) node {$G$};
  \path (g)+(-1,2) node[draw,shape=circle,fill] (u) {};
  \path (u)+(-0.5,0) node {$u$};
  \path (g)+(1,2) node[draw,shape=circle,fill] (u2) {};
  \path (u2)+(0.5,0) node {$u'$};
  \path (g)+(-1,3) node[draw,shape=circle,fill] (v) {};  
  \path (v)+(-0.5,0) node {$v$};
  \path (g)+(1,3) node[draw,shape=circle,fill] (v2) {}; 
  \path (v2)+(0.5,0) node {$w$};
 
  \draw (u2) -- (u) -- (v) -- (u2) (u) -- (v2) -- (u2)
(g)+(-2.9,-0.25) -- (u)
(g)+(2.8,-0.12) -- (u)
(g)+(-2.8,-0.12) -- (u2)
(g)+(2.9,-0.25) -- (u2);

  \draw[line width=1.2pt] (g)+(0,-0.5) ellipse (3cm and 1cm);

\end{tikzpicture}}}
\end{center}
\caption{Three reductions for diameter~$2$.}
\label{fig:red-diam2}
\end{figure}

Let $G$ be a graph. We define $f_1(G)$ to be the graph obtained from $G$ by adding a universal vertex $u$ and then, a neighbour $v$ of $u$ of degree~$1$. Similarly, $f_2(G)$ is the graph obtained from $f_1(G)$ by adding a twin $w$ of $v$. See Figures~\ref{fig:red-diam2-f1} and~\ref{fig:red-diam2-f2} for an illustration.

\begin{lem}\label{lemm:reduc-univ}
For any graph $G$, we have $\LD(f_1(G))=\LD(G)+1$. If $G$ is twin-free, $\ID(f_1(G))=\ID(G)+1$. If $G$ is open twin-free, $\OLD(f_2(G))=\OLD(G)+2$.
\end{lem}
\begin{proof}
Let $S$ be an identifying code of $G$. Then $S\cup\{v\}$ is also an identifying code of $f_1(G)$: all vertices within $V(G)$ are distinguished by $S$ as they were in $G$; vertex $v$ is dominated only by itself; vertex $u$ is the only vertex dominated by the whole set $S\cup\{v\}$. The same argument works for a locating-dominating set. Hence, $\LD(f_1(G))\leq\LD(G)+1$ and $\ID(f_1(G))\leq\ID(G)+1$. If $S$ is an open locating-dominating set of $G$, then similarly, $S\cup\{v,w\}$ is one of $f_2(G)$, hence $\OLD(f_2(G))\leq\OLD(G)+2$.

It remains to prove the converse. Let $S_1$ be an identifying code (or locating-dominating set) of
$f_1(G)$. Observe that $|S_1\cap\{u,v\}|\geq 1$ since $v$ must be
dominated. Hence if $S_1\setminus\{u,v\}$ is an identifying code (or
locating-dominating set) of $G$, we are done. Let us assume the
contrary. Then, necessarily $u\in S_1$ since $v$ does not dominate
any vertex of $V(G)$. But $u$ is a universal
vertex, hence $u$ does not separate any pair of vertices of
$V(G)$. Therefore, $S_1\setminus\{u\}$ separates all
pairs, but does not dominate some vertex $x\in V(G)$: we have $N[x]\cap S_1=\{u\}$. Note that $x$ is the only such vertex of $G$. This implies
that $v\in S_1$ (otherwise $x$ and $v$ are not separated by
$S_1$). But then $(S_1\setminus\{u,v\})\cup\{x\}$ is an identifying
code (or locating-dominating set) of $G$ of size $|S_1|-1$. This
completes the proof.

A similar proof works for open location-domination: if $S_2$ is an
open locating-dominating set of $f_2(G)$, then
$|S_2\cap\{u,v,w\}|\geq 2$ since $v,w$ must be separated and
totally dominated. Similarly, if $S_2\setminus\{u,v,w\}$ is an
open locating-dominating set of $G$, we are done. Otherwise, again $u$
must belong to $S_2$, and is needed only for domination. But then if
there is a vertex among $v,w$ that is not in $S_2$, the other one
would not be separated from the vertex $x$ only dominated by $u$. But
then $S_2\setminus\{u,v,w\}\cup\{y\}$, for any vertex $y\in N(x)$, is an open
locating-dominating set of size $|S_2|-2$ and we are done.
\end{proof}

Lemma~\ref{lemm:reduc-univ} directly implies the following theorem:

\begin{thm}\label{thm:reduc-univ}
Let $\mathcal C$ be a class of graphs that is closed under the graph transformation $f_1$ ($f_2$, respectively). If \PBID{} or \PBLD{} (\PBOLD, respectively) is NP-complete for graphs in $\mathcal C$, then it is also NP-complete for graphs in $\mathcal C$ that have diameter~$2$.
\end{thm}

Theorem~\ref{thm:reduc-univ} can be applied to the classes of split graphs (for $f_1$), interval graphs and permutation graphs (for both $f_1$ and $f_2$). By the results about split graphs from~\cite{F13j} and about interval graphs and permutation graphs of Corollary~\ref{coro:perm-inter-NPc}, we have:

\begin{cor}
\PBID{} and \PBLD{} are NP-complete for split graphs of diameter~$2$. \PBID, \PBLD{} and \PBOLD{} are NP-complete for interval graphs of diameter~$2$ and for permutation graphs of diameter~$2$.
\end{cor}

We now a give a similar reduction from \PBLD{} to \PBMD. Given a graph $G$, let $f_3(G)$ be the graph obtained from $G$ by adding two adjacent universal vertices $u,u'$ and then, two non-adjacent vertices $v$ and $w$ that are only adjacent to $u$ and $u'$ (see Figure~\ref{fig:red-diam2-f3} for an illustration).

\begin{lem}\label{lemm:reduc-univ-MD}
For any graph $G$, $\MD(f_3(G))=\LD(G)+2$.
\end{lem}
\begin{proof}
Let $S$ be a locating-dominating set of $G$. We claim that
$S_3=S\cup\{u,v\}$ is a resolving set of $f_3(G)$. Every vertex of $S_3$
is clearly distinguished. Every original vertex of $G$ is determined
by a distinct set of vertices of $S$ that are at distance~$1$ of
it. Vertex $u'$ is the only vertex to be at distance~$1$ of each vertex
in $S_3$. Finally, vertex $w$ is the only vertex to be at distance~$1$
of $u$ and at distance~$2$ from all other vertices of $S_3$.

For the other direction, assume $B$ is a resolving set of $f_3(G)$. Then
necessarily one of $u,u'$ (say $u$) belongs to $B$; similarly, one of
$v,w$ (say $v$) belongs to $B$. Hence, if the restriction
$B_G=B\cap V(G)$ is a locating-dominating set of $G$, we are
done. Otherwise, since no vertex among $u,u',v,w$ may distinguish
any pair of $G$ and since vertices of $G$ are at distance at most $2$ in $f_3(G)$, all the sets $N[x]\cap B$ are distinct for $x\in V(G)\setminus B_G$. But $B_G$ is not a locating-dominating set, so there is a (unique) $x$ vertex of $G$ that is
not dominated by $B_G$ in $G$. If $|B\cap\{u,u',v,w\}|\geq 3$,
$B_G\cup\{x\}$ is a locating-dominating set of size at most $|B|-2$
and we are done. Otherwise, note that in $f_3(G)$, $x$ is at distance~$1$
from $u$ and at distance~$2$ from all other vertices of $B$. But this is
also the case for $w$, which is not separated from $x$ by $B$, which is a
contradiction.
\end{proof}

We obtain the following results:

\begin{thm}\label{thm:reduc-univ-MD}
Let $\mathcal C$ be a class of graphs that is closed under the graph transformation $f_3$. If \PBLD{} is NP-complete for graphs in $\mathcal C$, then \PBMD{} is also NP-complete for graphs in $\mathcal C$ that have diameter~$2$.
\end{thm}

Again, using the results about split graphs from~\cite{F13j} and about interval graphs and permutation graphs of Corollary~\ref{coro:perm-inter-NPc}, we have:

\begin{cor}
\PBMD{} is NP-complete for split graphs of diameter~$2$, for interval graphs of diameter~$2$ and for permutation graphs of diameter~$2$.
\end{cor}

\section{\PBMD{} parameterized by solution size is FPT on interval graphs}\label{sec:MD-interval-FPT}

The purpose of this section is to prove that \PBMD{} (parameterized by solution size) is FPT on interval graphs. We begin with preliminary results, before describing our algorithm and proving its correctness. The algorithm is based on dynamic programming over a path-decomposition.

\subsection{Preliminaries}

We start by stating a few properties and lemmas that are necessary for our algorithm.

\subsubsection{Interval graphs}

Given an interval graph $G$, we can assume that in its interval model, all endpoints are distinct, and that the intervals are closed intervals. Given an interval $I$, we will denote by $\ell(I)$ and by $r(I)$ its left and right endpoints, respectively. We define two natural total orderings of $V(G)$ based on this model: $x <_L y$ if and only if the left endpoint of $x$ is smaller then the left endpoint of $y$, and $x <_R y$ if and only if the right endpoint of $x$ is smaller than the right endpoint of $y$.

Given a graph $G$, its \emph{distance-power} $G^d$ is the graph obtained from $G$ by adding an edge between each pair of vertices at distance at most~$d$ in $G$. We will use the following result.

\begin{thm}[\cite{ADH}]\label{prop:order-in-power}
Let $G$ be an interval graph with an interval model inducing orders $<_L$ and $<_R$, and let $d\geq 2$ be an integer. Then the power graph $G^d$ is an interval graph with an interval model inducing the same orders $<_L$ and $<_R$ as $G$ (that can be computed in linear time).
\end{thm}

\subsubsection{Tree-decompositions} 

\begin{defi}\label{def:treedec}
A \emph{tree-decomposition} of a graph $G$ is a pair $(\mathscr{T},\mathcal{X})$, where $\mathscr{T}$ is a tree and $\mathcal{X}:=\{X_t:t\in V(\mathscr{T})\}$ is a collection of subsets of $V(G)$ (called \emph{bags}), such that they satisfy the following conditions:\\
(i)~$\bigcup_{t\in V(\mathscr{T})} X_t=V(G)$;\\
(ii)~for every edge $uv\in E(G)$, there is a bag of $\mathcal{X}$ that contains both $u$ and $v$;\\
(iii)~for every vertex $v\in V(G)$, the set of bags containing $v$ induces a subtree of $\mathscr{T}$.
\end{defi}

Given a tree-decomposition of $(\mathscr{T},\mathcal{X})$, the maximum size of a bag $X_t$ over all tree nodes $t$ of $\mathscr{T}$ minus one is called the \emph{width} of $(\mathscr{T},\mathcal{X})$. The minimum width of a tree-decomposition of $G$ is the \emph{treewidth} of $G$. The notion of tree-decomposition has been used extensively in algorithm design, especially via dynamic programming over the tree-decomposition.

We consider a {\em rooted} tree-decomposition by fixing a root of $\mathscr{T}$ and orienting the tree edges from the root toward the leaves. A rooted tree-decomposition is \emph{nice} (see Kloks~\cite{niceTW}) if each node $t$ of $\mathscr{T}$ has at most two children and falls into one of the four types:\\
(i) {\em Join} node: $t$ has exactly two children $t_1$ and $t_2$, and $X_t=X_{t_1}=X_{t_2}$.\\
(ii) {\em Introduce} node: $t$ has a unique child $t'$, and $X_t=X_{t'}\cup\{v\}$.\\
(iii) {\em Forget} node: $t$ has a unique child $t'$, and $X_t=X_{t'}\setminus \{v\}$.\\
(iv) {\em Leaf} node: $t$ is a leaf node in $\mathscr{T}$. 

Given a tree-decomposition, a nice tree-decomposition of the same width always exists and can be computed in linear time~\cite{niceTW}.

If $G$ is an interval graph, we can construct a tree-decomposition of $G$ (in fact, a path-decomposition) with special properties.

\begin{prop}\label{prop:path-dec}
Let $G$ be an interval graph with clique number $\omega$ and an interval model inducing orders $<_L$ and $<_R$. Then, $G$ has a nice tree-decomposition $(\mathscr{P},\mathcal{X})$ of width $\omega-1$ that can be computed in linear time, where moreover:\\
(a) $\mathscr{P}$ is a path (hence there are no join nodes);\\
(b) every bag is a clique;\\
(c) going through $\mathscr{P}$ from the leaf to the root, the order in which vertices are introduced in an introduce node corresponds to $<_L$;\\
(d) going through $\mathscr{P}$ from the leaf to the root, the order in which vertices are forgotten in a forget node corresponds to $<_R$;\\
(e) the root's bag is empty, and the leaf's bag contains only one vertex.
\end{prop}
\begin{proof}
Given a graph $G$, one can decide if it is an interval graph and, if so, compute a representation of it in linear time~\cite{BL76}. This also gives us the ordered set of endpoints of intervals of $G$.

To obtain $(\mathscr{P},\mathcal{X})$, we first create the leaf node $t$, whose bag $X_t$ contains the interval with smallest left endpoint. We then go through the set of all endpoints of intervals of $G$, from the second smallest to the largest. Let $t$ be the last created node. If the new endpoint is a left endpoint $\ell(I)$, we create an introduce node $t'$ with $X_{t'}=X_t\cup \{I\}$. If the new endpoint is a right endpoint $r(I)$, we create a forget node $t'$ with $X_{t'}=X_t\setminus \{I\}$. In the end we create the root node as a forget node $t$ with $X_t=\emptyset$ that forgets the last interval of $G$.

Observe that one can associate to every node $t$ (except the root) a point $p$ of the real line, such that the bag $X_t$ contains precisely the set of intervals containing $p$: if $t$ is an introduce node, $p$ is the point $\ell(I)$ associated to the creation of $t$, and if $t$ is a forget node, it is the point $r(I)+\epsilon$, where $\epsilon$ is sufficiently small and $r(I)$ is the endpoint associated to the creation of $t$. This set forms a clique, proving Property~(b). Furthermore this implies that the maximum size of a bag is $\omega$, hence the width is at most $\omega-1$ (and at least $\omega-1$ since every clique must be included in some bag).
 
Moreover it is clear that the procedure is linear-time, and by construction, Properties~(a), (c), (d), (e) are fulfilled.

Let us now show that $(\mathscr{P},\mathcal{X})$ is a tree-decomposition. It is clear that every vertex belongs to some bag, proving Property~(i) of Definition~\ref{def:treedec}. Moreover let $u,v$ be two adjacent vertices of $G$, and assume $u<_L v$. Then, consider the introduce node of $\mathscr{P}$ where $v$ is introduced. Since $u$ has started before $v$ but has not stopped before the start of $v$, both $u,v$ belong to $X_t$, proving Property~(ii). Finally, note that a vertex $v$ appears exactly in all bags starting from the bag where $v$ is introduced, until the bag where $v$ is forgotten. Hence Property~(iii) is fulfilled, and the proof is complete.
\end{proof}

The following lemma immediately follows from Theorem~\ref{prop:order-in-power}.

\begin{lem}\label{lemm:N_4-intro-forget}
Let $G$ be an interval graph with an interval model inducing orders $<_L$ and $<_R$, let $d\geq 1$ be an integer and let $(\mathscr{P},\mathcal{X})$ be a tree-decomposition of $G^d$ obtained by Proposition~\ref{prop:path-dec} (recall that by Theorem~\ref{prop:order-in-power}, $G^d$ is an interval graph, and it has an intersection model inducing the same orders $<_L$ and $<_R$). Then the following holds.\\
(a) Let $t$ be an introduce node of $(\mathscr{P},\mathcal{X})$ with child $t'$, with $X_t=X_{t'}\cup\{v\}$. Then, $X_{t}$ contains every vertex $w$ in $G$ such that $d_G(v,w)\leq d$ and $w<_L v$.\\
(b) Let $t'$ be the child of a forget node $t$ of $(\mathscr{P},\mathcal{X})$, with $X_t=X_{t'}\setminus\{v\}$. Then, $X_{t'}$ contains every vertex $w$ in $G$ such that $d_G(v,w)\leq d$ and $v<_R w$.
\end{lem}
\begin{proof}
We prove~(a), the proof of~(b) is the same. By Theorem~\ref{prop:order-in-power}, we may assume that $<_L$ is the same in $G$ and $G^d$. By construction of $(\mathscr{P},\mathcal{X})$ the introduce node of $v$ contains all intervals $w$ of $G^d$ intersecting $v$ with $w<_L v$ in $G^d$. Hence $w<_L v$ in $G$ as well, and $d_G(v,w)\leq d$.
\end{proof}

\subsubsection{Lemmas for the algorithm}

We now prove a few preliminary results necessary for the argumentation. We first start with a definition and a series of lemmas based on the linear structure of an interval graph, that will enable us to defer the decision-taking (about which vertex should belong to the solution in order distinguish a specific vertex pair) to later steps of the dynamic programming.

\begin{defi}\label{def:leftmost-rightmost-paths}
Given a vertex $u$ of an interval graph $G$, the \emph{rightmost path} $P_R(u)$ of $u$ is the path $u^R_0,\ldots,u^R_p$ where $u=u^R_0$, for every $u^R_i$ ($i\in\{0,\ldots,p-1\}$) $u^R_{i+1}$ is the neighbour of $u^R_i$ with the largest right endpoint, and thus $u^R_p$ is the interval in $G$ with largest right endpoint. Similarly, we define the \emph{leftmost path} $P_L(u)=u^L_0,\ldots,u^L_q$ where for every $u^L_i$ ($i\in\{0,\ldots,q-1\}$) $u^L_{i+1}$ is the neighbour of $u^L_i$ with the \emph{smallest} \emph{left} endpoint.
\end{defi}

Note that $P_R(u)$ and $P_L(u)$ are two shortest paths from $u^R_0$ to $u^R_p$ and $u^L_q$, respectively.

\begin{lem}\label{lemma:comput-distances-right-leftmost-paths}
Let $u$ be an interval in an interval graph $G$ and $P_R(u)=u_0^R,\ldots,u_p^R$ be the rightmost path of $u$, and let $v$ be an interval starting after the end of $u^{R}_{i-1}$ ($i\in\{1,\ldots,p\}$), where $u^{R}_{i-1}\in P_R(u)$. Then $d(u,v)=d(u^R_i,v)+i$. Similarly, if $v$ ends before the start of an interval $u^{L}_{i-1}$ in $P_L(u)=u^L_0,\ldots,u^L_q$ ($i\in\{1,\ldots,q\}$), then $d(u,v)=d(u^L_i,v)+i$. 
\end{lem}
\begin{proof}
We prove the claim only for the first case, the second one is symmetric. 
Consider the shortest path from $u$ to $v$ by choosing the interval intersecting $u$ that has the largest right endpoint, and iterating. This path coincides with $P_R(u)$ until it contains some interval $u^R_j$ such that $u^R_j$ intersects $v$. Since $v$ starts after the end of $u^{R}_{i-1}$, we have $i\leq j$. Thus, the interval $u^R_i$ lies on a shortest path from $u$ to $v$, and hence $d(u,v)=d(u^R_i,v)+d(u,u^R_i)=d(u^R_i,v)+i$.
\end{proof}

\begin{lem}\label{lemma:dist-u_i-v_i}
Let $u,v$ be a pair of intervals of an interval graph $G$ and $P_R(u)=u_0^R,\ldots,u_p^R$, $P_R(v)=v_0^R,\ldots,v_{p'}^R$ their corresponding rightmost paths (recall that $u_p^R=v_{p'}^R$). Assuming that $p\leq p'$, for every $u^R_i\in P_R(u)$ and $v^R_i\in P_R(v)$ such that $i\in\{0,\ldots,p\}$, we have $d(u^R_i,v^R_i)\leq d(u,v)$.
\end{lem}
\begin{proof}
First note that, by letting $w=u^R_i$, we have $w^R_1=u^R_{i+1}$. Therefore, we only need to prove the claim for $i=1$.

If $u$ and $v$ are adjacent, then either $v=u^R_1$ (then we are done) or $u^R_1$ must end after $v$. Then, either $u^R_1$ intersects $v^R_1$, or $u^R_1=v^R_1$. In both cases, $d(u^R_1,v^R_1)\leq 1$.

If $u$ and $v$ are not adjacent, we can assume that $u$ ends before $v$ starts. Then, by Lemma~\ref{lemma:comput-distances-right-leftmost-paths}, $d(u^R_1,v)=d(u,v)-1$ and $d(u^R_1,v^R_1)\leq d(u^R_1,v)+d(v,v^R_1) =d(u,v)-1+1=d(u,v)$.
\end{proof}

We say that a pair $u,v$ of intervals in an interval graph $G$ is separated by interval $x$ \emph{strictly from the right} (\emph{strictly from the left}, respectively) if $x$ starts after both right endpoints of $u,v$ (ends before both left endpoints of $u,v$ respectively). In other words, $x$ is not a neighbour of any of $u$ and $v$.

The next lemma is crucial for our algorithm.

\begin{lem}\label{lemma:equiv-distances-right-leftmost-paths}
Let $u,v,x$ be three intervals in an interval graph $G$ and let $i$ be an integer such that $x$ starts after both right endpoints of $u^R_i\in P_R(u)$ and $v^R_i\in P_R(v)$. Then the three following facts are equivalent:\\
(1) $x$ separates $u^R_i$, $v^R_i$;\\
(2) for every $j$ with $0\leq j\leq i$, $x$ separates $u^R_j$, $v^R_j$;\\
(3) for some $j$ with $0\leq j \leq i$, $x$ separates $u^R_j$, $v^R_j$.\\
Similarly, assume that $x$ ends before both left endpoints of $u^L_i\in P_L(u)$ and $v^L_i\in P_L(v)$. Then the three following facts are equivalent:\\
(i) $x$ separates $u^L_i$, $v^L_i$;\\
(ii) for every $j$ with $0\leq j\leq i$, $x$ separates $u^L_j$, $v^L_j$;\\
(iii) for some $j$ with $0\leq j\leq i$, $x$ separates $u^L_j$, $v^L_j$.\\
\end{lem}
\begin{proof}
We prove only (1)--(3), the proof of (i)--(iii) is symmetric. Let $0\leq j\leq i$ and $u'=u^R_j$ and $v'=v^R_j$. Then $(u')^R_{i-j}=u^R_i$ and $(v')^R_{i-j}=v^R_i$. By Lemma~\ref{lemma:comput-distances-right-leftmost-paths}, $d(u^R_j,x)=d(u^R_i,x)+(j-i)$ and similarly $d(v^R_j,x)=d(v^R_i,x)+(j-i)$. 
Hence $x$ separates $u^R_i$ and $v^R_i$ if and only if it separates $u^R_j$ and $v^R_j$ which implies the lemma.
\end{proof}

We now introduce a local version of resolving sets that will be used in our algorithm.

\begin{defi}
A \emph{distance-$2$ resolving set} is a set $S$ of vertices where for each pair $u,v$ of vertices at distance at most~$2$, there is a vertex $x\in S$ such that $d(u,s)\neq d(v,s)$.
\end{defi}

Using the following lemma, we can manage to ``localize'' the dynamic programming, as we will only need to distinguish pairs of vertices that will be present together in one bag.

\begin{lem}\label{lemma:dist-2-resolv}
Any distance-$2$ resolving set of an interval graph $G$ is a resolving set of $G$.
\end{lem}
\begin{proof}
Assume to the contrary that $S$ is a distance-$2$ resolving set of an interval graph $G$ but not a resolving set. It means that there is a pair of vertices $u,v$ at distance at least~$3$ that are not separated by any vertex of $S$. Among all such pairs, we choose one, say $\{u,v\}$, such that $d(u,v)$ is minimized. Without loss of generality, we assume that $u$ ends before $v$ starts. 

Consider $u^R_1$ ($v^L_1$, respectively), the interval intersecting $u$ ($v$, respectively) that has the largest right endpoint (smallest left endpoint, respectively). We have $u^R_1\neq v^L_1$ (since $d(u,v)\geq 3$) and $d(u^R_1,v^L_1)=d(u,v)-2<d(u,v)$. By minimality, $u^R_1$ and $v^L_1$ are separated by some vertex $s\in S$. But $s$ does not separate $u$ and $v$, thus $s\notin \{u^R_1,v^L_1\}$.

Without loss of generality, we can assume that $d(u^R_1,s)<d(v^L_1,s)$. In particular, $d(v^L_1,s)\geq 2$ and $s$ is ending before $v^L_1$ starts. Thus, by Lemma~\ref{lemma:comput-distances-right-leftmost-paths}, $d(v,s)=d(v^L_1,s)+1$. However, we also have $d(u,s)\leq d(u^R_1,s)+1 \leq d(v^L_1,s)<d(v,s)$. Hence $s$ is separating $u$ and $v$, a contradiction.
\end{proof}

The next lemma, which is a slightly modified version of a result in our paper~\cite{part1}, enables us to upper-bound the size of the bags in our tree-decompositions, which will induce diameter~$4$-subgraphs of $G$.

\begin{lem}\label{lemma:bnded-diameter-subgraph}
Let $G$ be an interval graph with a resolving set of size~$k$, and let $B\subseteq V(G)$ be a subset of vertices such that for each pair $u,v\in B$, $d_G(u,v)\leq d$. Then $|B|\leq 4dk^2+(2d+3)k+1$.
\end{lem}
\begin{proof}
Let $s_1,\ldots,s_k$ be the elements of a resolving set $S$ of size~$k$ in $G$. Consider an interval representation of $G$, and let $\mathcal B$ be the minimal segment of the real line containing all intervals corresponding to vertices of $B$.

For each $i$ in $\{1,\ldots,k\}$, consider the leftmost and rightmost paths $P_L(s_i)$ and $P_R(s_i)$, as defined in Definition~\ref{def:leftmost-rightmost-paths}. Let $L^i$ be the ordered set of left endpoints of intervals of $P_L(s_i)$, and let $R^i$ be the ordered set of right endpoints of intervals of $P_R(s_i)$. Note that intervals at distance~$j$ of $s_i$ in $G$ are exactly the intervals finishing between $\ell(u^L_{j+1})$ and $\ell(u^L_{j})$, or starting between $r(u^R_{j})$ and $r(u^R_{j+1})$. Hence, for any interval of $G$, its distance to $s_i$ is uniquely determined by the position of its right endpoint in the ordered set $L^i$ and the position of its left endpoint in the ordered set $R^i$. Moreover, note that, since any two vertices in $B$ are at distance at most~$d$, $\mathcal B$ may contain at most~$d$ points of $L^i$ and at most~$d$ points of $R^i$.

Therefore, $\mathcal B$ may contain at most~$2kd$ points of $\bigcup_{1\leq i\leq k} (L^i\cup R^i)$. This set of points defines a natural partition $\mathcal P$ of $\mathcal B$ into at most $2kd+1$ sub-segments, and any interval of $B$ is uniquely determined by the positions of its two endpoints in $\mathcal P$ (if two intervals start and end in the same part of $\mathcal P$, they are not separated by $S$, a contradiction).

Let $I\in B\setminus S$. For a fixed $i\in\{1,\ldots,k\}$, by definition of the sets $L^i$, the interval $I$ cannot contain two points of $L^i$ and similarly, it cannot contain two points of $R^i$. Thus, $I$ contains at most $2k$ points of the union of all the sets $L^i$ and $R^i$. Therefore, if $P$ denotes a part of $\mathcal P$, there are at most $2k+1$ intervals with left endpoints in $P$. In total, there are at most $(2kd+1)\cdot (2k+1)$ intervals in $B\setminus S$ and hence $|B|\leq (2kd+1)\cdot (2k+1)+k=4dk^2+(2d+3)k+1$.
\end{proof}

\subsection{The algorithm}

We are now ready to describe our algorithm.

\begin{thm}\label{thm:MD-interval-FPT}
\PBMD{} can be solved in time $2^{O(k^4)}n$ on interval graphs, i.e. it is FPT on this class when parameterized by the solution size $k$.
\end{thm}
\begin{proof}
Let $(\mathscr{P},\mathcal{X})$ be a path-decomposition of $G^4$ (which by Theorem~\ref{prop:order-in-power} is an interval graph) obtained using Proposition~\ref{prop:path-dec}.

The algorithm is a bottom-up dynamic programming on $(\mathscr{P},\mathcal{X})$. By Proposition~\ref{prop:path-dec}(b), every bag of $(\mathscr{P},\mathcal{X})$ is a clique of $G^4$ (i.e. an induced subgraph of diameter at most~$4$ in $G$) and hence by Lemma~\ref{lemma:bnded-diameter-subgraph}, it has $O(k^2)$ vertices. Thanks to Lemma~\ref{lemma:equiv-distances-right-leftmost-paths}, we can ``localize'' the problem by considering for separation, only pairs of vertices present together in the current bag. Let us now be more precise.

For a node $t$ in $\mathscr{P}$, we denote by $\mathcal P(X_t)$ the pairs of intervals in $X_t$ that are at distance at most~$2$ (in $G$).

For each node $t$, we compute a set of \emph{configurations} using the configurations of the child of $t$ in $\mathscr{P}$. A configuration contains full information about the local solution on $X_t$, but also stores necessary information about the vertex pairs that still need to be separated. 
More precisely, a configuration $C=(\mS,\msep,\msepr,\mcnt)$ of $t$ is a tuple where:

\begin{itemize}
\item $\mS\subseteq X_t$ contains the vertices of the sought solution belonging to $X_t$;
\item $\msep: \mathcal P(X_t)\to \{0,1,2\}$ assigns, to every pair in $\mathcal P(X_t)$, value~0 if the pair has not yet been separated, value~$2$ if it has been separated strictly from the left, and value~$1$ otherwise;
\item $\msepr:\mathcal P(X_t)\to \{0,1\}$ assigns, to every pair in $\mathcal P(X_t)$, value~$1$ if the pair needs to be separated strictly from the right (and it is not yet separated), and value~0 otherwise; 
\item $\mcnt$ is an integer counting the total number of vertices in the partial solution that has led to $C$.
\end{itemize}

Starting with the leaf of $\mathscr{P}$, for each node our algorithm goes through all possibilities of choosing $\mS$; however, $\msep$, $\msepr$ and $\mcnt$ are computed along the way. At each new visited node $t$ of $\mathscr{P}$, a set of configurations is constructed from the configuration sets of the child of $t$. The algorithm makes sure that all the information is consistent, and that configurations that will not lead to a valid resolving set (or with $\mcnt>k$) are discarded.

\smallskip

\noindent {\bf Leaf node:} For the leaf node $t$, since by Proposition~\ref{prop:path-dec}(e) $X_t=\{v\}$, we create two configurations $C_1=(\emptyset,\msep,\msepr,0)$ and $C_2=(\{v\},\msep,\msepr,1)$ (where $\msep$ and $\msepr$ are empty in both configurations).

\medskip

\noindent {\bf Introduce node:} Let $t$ be an introduce node with $t'$ its child, where $X_t=X_{t'}\cup \{v\}$. For every configuration $(\mS',\msep',\msepr',\mcnt')$ of $t'$, we create two configurations $C_1=(\mS'\cup\{v\},\msep_1,\msepr_1,\mcnt'+1)$ (corresponding to the case where $v$ is in the partial solution) and $C_2=(\mS',\msep_2,\msepr_2,\mcnt')$ (where $v$ is not added to the partial solution).

The elements of $\msep_1$ and $\msepr_1$ in $C_1$ are first copied from $\msep'$ and $\msepr'$, and updated by checking, for every pair $x,y$ of $\mathcal P(X_t)$ whether $v$ separates $x,y$ (note that $v$ cannot separate any such pair strictly from the left). Also note that $v$ is separated from all other vertices since it belongs to the solution, but for $x=v$ we still need to check whether $v,y$ are strictly separated from the left (in which case we set $\msep_1(v,y)=2$, otherwise $\msep_1(v,y)=1$). To do this, we compute $v^L_1$ and $y^L_1$ (by Lemma~\ref{lemm:N_4-intro-forget}(a) they both belong to $X_t$), and we first check if they are strictly separated from the left, which is true if and only if $\msep'(v^L_1,y^L_1)=2$. If $v^L_1$ and $y^L_1$ are separated strictly from the left, then so are $v$ and $y$. Otherwise, if $v$ and $y$ are still strictly separated from the left, there must be an interval $z$ ending before the left endpoint of $y$ and separating $v,y$. Since $z$ does not separate $v^L_1$ and $y^L_1$ strictly from the left, $z$ must be adjacent to $y^L_1$ and thus $d_G(v,z)\leq 4$ (since $d_G(v,y)\leq 2$). Then, by Lemma~\ref{lemm:N_4-intro-forget}, $z$ belongs to $X_t$, thus it is enough to test whether any vertex of $\mS'$ separates $v,y$ strictly from the left. Moreover, we let $\msepr_1(v,y)=0$.

For $C_2$, we must compute $\msep_2(v,w)$ and $\msepr_2(v,w)$ for every $w$ such that $(v,w)\in\mathcal P(X_t)$. To do so, we consider the first intervals of $P_L(v)$ and $P_L(w)$. We let $\msep_2(v,w)=2$ if for the pair $v^L_1,w^L_1$ with $v^L_1\in P_L(v)$ and $w^L_1\in P_L(w)$, $\msep'(v^L_1,w^L_1)=2$, or if some vertex of $\mS'$ separates $v,w$ strictly from the left. Otherwise, if $v,w$ are separated by a neighbour of $w$, we set $\msep_2(v,w)=1$. We also compute $\msepr_2$ from $\msepr'$ by letting $\msepr_2(v,w)=0$ and copying all other values.

If $\mcnt+1>k$, $C_1$ is discarded. The remaining valid configurations among $C_1,C_2$ are added to the set of configurations of $t$. If in this set, there are two configurations that differ only on their value of $\mcnt$, we only keep the one with the smallest value of $\mcnt$.

\medskip

\noindent {\bf Forget node:} Let $t$ be a forget node and $t'$ be its child, with $X_t=X_{t'}\setminus\{v\}$. For every configuration $(\mS',\msep',\msepr',\mcnt')$ of $t'$, we create the configuration $(\mS'\setminus\{v\},\msep,\msepr,\mcnt')$. We create $\msep$ and $\msepr$ by copying all entries $\msep'(x,y)$ and $\msepr'(x,y)$ such that $x,y\in \mathcal P(X_t)$.

For every vertex $w$ in $X_t$ such that $d_G(v,w)\leq 2$, if $\msep'(v,w)=0$ or $\msepr'(v,w)=1$ (i.e. $v,w$ still need to be separated strictly from the right), we determine $v^R_1$ and $w^R_1$ and let $\msepr(v^R_1,w^R_1)=1$ (note that $d_G(v,v^R_1)=1$, $d_G(v,w^R_1)\leq 3$, $v<_R v^R_1$ and $v<_R w^R_1$, hence by Lemma~\ref{lemm:N_4-intro-forget}(b) $v^R_1,w^R_1\in X_{t'}$ and hence $v^R_1,w^R_1\in X_{t}$). However, if $v^R_1=w^R_1$, we discard the current configuration. Indeed, by Lemma~\ref{lemma:equiv-distances-right-leftmost-paths}, $v,w$ cannot be separated strictly from the right: any shortest path to any of $v,w$ from some vertex $x$ whose interval starts after both right endpoints of $v,w$ must go through $v^R_1=w^R_1$ and hence $d(x,v^R_1)=d(x,w^R_1)$. We also discard the configuration if $v^R_1$ or $w^R_1$ does not exist (i.e. $v$ or $w$ is the rightmost interval of $G$).

Finally, if there are two configurations that differ only on their value of $\mcnt$, again we only keep the one with the smallest value of $\mcnt$.

\medskip

\noindent {\bf Root node:} At root node $t$, since by Proposition~\ref{prop:path-dec}(e) $X_t=\emptyset$, $t$ has at most one configuration. We output ``yes'' only if this configuration exists, and if $\mcnt\leq k$. Otherwise, we output ``no''.

\medskip

We now analyze the algorithm.

\smallskip

\noindent\textbf{Correctness.} We claim that $G$ has a resolving set of size at most~$k$ if and only if the root node of $\mathscr{P}$ contains a valid configuration. By Lemma~\ref{lemma:dist-2-resolv}, this is equivalent to proving that $G$ has an optimal \emph{distance-$2$ resolving set} of size at most~$k$ if and only if the root node of $\mathscr{P}$ contains a valid configuration.
First, assume that the dynamic programming has succeeded, i.e. the root bag contains a valid configuration $C$. Assume that $C$ has smallest value $\mcnt$. We want to prove that the union of all partial solutions $\mS$ of all configurations that have led to the computation of $C$ is a valid optimal solution $S$.

\smallskip

We first prove that for every pair $u,v$ of vertices with $d_G(u,v)\leq 2$ and $u<_R v$, $S$ separates $u,v$. By Lemma~\ref{lemm:N_4-intro-forget}(b), $u,v$ are present together in the child $t'$ of forget node $t$ of $\mathscr{P}$ where $u$ is forgotten. Let $C_{t'}=(\mS',\msep',\msepr',\mcnt')$ and $C_t=(\mS,\msep,\msepr,\mcnt)$ be the configurations of $t',t$ that have led to the end configuration $C$. In the computation of $C_t$, since $C_t$ was not discarded, we either had $\msep'(u,v)>0$ in $C_{t'}$ or the algorithm has set $\msepr(u^1_r,v^1_r)=1$, in which case $u^R_1\neq v^R_1$. Assume we had $\msep'(u,v)=1$. Then, in some configuration $C_{t''}$ that has led to computing $C_{t'}$ (possibly $t'=t''$), $u$ and $v$ were separated by some vertex in $S$ belonging to $C_{t''}$, and we are done. If $\msep'(u,v)=2$, similarly either $u,v$ have been separated by some vertex of $S$ belonging to a (possibly earlier) configuration, or we had $\msep(u^L_i,v^L_i)=2$, in which case by Lemma~\ref{lemma:equiv-distances-right-leftmost-paths} we are also done. If however, the algorithm has set $\msepr(u^R_1,v^R_1)=1$, recall that unless in some bag $u^R_1,v^R_1$ is separated strictly from the right, when we forget $u^R_1$ we set $\msepr(u^R_2,v^R_2)=1$. Hence, since $C$ was a valid configuration (and has not been discarded), at some step we have separated $u^R_i,v^R_i$ strictly from the right, which by Lemma~\ref{lemma:equiv-distances-right-leftmost-paths} implies that $u,v$ are separated by $S$, and we are done.

Moreover $S$ is optimal because we have chosen $C$ so as to minimize the size $\mcnt$ of the overall solution. At each step, the algorithm discards, among equivalent configurations, the ones with larger values of $\mcnt$, ensuring that the size of the solution is minimized. This proves our claim.

\smallskip

For the converse, assume that $G$ has an optimal distance-$2$ resolving set $S$ of size at most~$k$. We will need the following claim.

\begin{claimbis}\label{claim:sep-by-neighbour}
Let $u,v$ be a pair of vertices with $d_G(u,v)\leq 2$. Then, any vertex $x$ that could separate $u,v$ neither strictly from the right nor strictly from the left is present in some bag together with both $u,v$.
\end{claimbis}
\begin{proofofclaim}
Necessarily, $x$ is a neighbour of one of $u,v$ in $G$. Hence $d_G(x,u)\leq 3$ and $d_G(x,v)\leq 3$. If $x<_L v$, by Lemma~\ref{lemm:N_4-intro-forget}(a) $x,u,v$ are present in the bag where $v$ is introduced. If $v<_L x$, similarly $x,u,v$ are present in the bag where $x$ is introduced.
\end{proofofclaim}

We will prove that some configuration $C$ was computed using a series of configurations where for each node $t$ of $\mathscr{P}$, the right subset $S\cap X_t$ was guessed. By contradiction, if this was not the case, then at some step of the algorithm we would have discarded a configuration $C'$ although it arised from guessing the correct partial solution of $S$. Since $S$ is optimal, $C'$ was not discarded because there was a copy of $C'$ with different value of counter $\mcnt$ (otherwise this copy would lead to a solution strictly smaller than $S$). Hence the discarding of $C'$ has happened at a node $t$ that is a forget node. Assume that $t$ is a forget node where vertex $v$ was forgotten (assume $t'$ is the child of $t$ in $\mathscr{P}$). This happens only if for some $w\in X_t$ with $d_G(v,w)\leq 2$, we had either (i)~$\msep'(v,w)=0$ and $v^R_1=w^R_1$, or (ii)~$\msepr(v,w)=1$ and $v^R_1=w^R_1$. If (i) holds, then $v,w$ are considered not to be separated, although they are actually separated (by our assumption on $C'$). Since $v^R_1=w^R_1$, $v^R_1$ and $w^R_1$ cannot be separated strictly from the right, hence by Lemma~\ref{lemma:equiv-distances-right-leftmost-paths} $v,w$ are not separated strictly from the right. If they are not separated strictly from the left, Claim~\ref{claim:sep-by-neighbour} implies a contradiction because the vertex separating $v,w$ was present together in a bag with $v,w$ and hence we must have $\msep'(v,w)=1$. Hence, $v,w$ are separated strictly from the left. But again by Lemma~\ref{lemma:equiv-distances-right-leftmost-paths}, this means that some vertices $v^L_i, w^L_i$ in $P_R(v)\times P_R(w)$ have been separated strictly from the left (assume that $i$ is maximal with this property). Since by Lemma~\ref{lemma:dist-u_i-v_i}, $d_G(v^L_i,w^L_i)\leq 2$, by Lemma~\ref{lemm:N_4-intro-forget} these two vertices were present in some bag simultaneously, together with the vertex that is strictly separating them from the left (and has distance at most~$4$ from $w^L_i$). Then in the configuration corresponding to this bag, $\msep(v^L_i,w^L_i)=2$, and we had $\msep'(v,w)=2$ in $C'$, a contradiction. If (ii) holds, there exists a pair $x,y$ such that in some earlier configuration, we had $\msep(x,y)=0$, $v=x^R_i\in P_R(x)$ and $w=y^R_i\in P_R(y)$. By the same reasoning as for (i) we obtain a contradiction. This proves this side of the implication, and completes the proof of correctness.

\smallskip

\noindent\textbf{Running time.} At each step of the dynamic programming, we compute the configurations of a bag from the set of configurations of the child bag. The computation of each configuration is polynomial in the size of the current bag of $(\mathscr{P},\mathcal{X})$. Since a configuration is precisely determined by a tuple $(\mS,\msep,\msepr)$ (if there are two configurations where only $\mcnt$ differs, we only keep the one with smallest value), there are at most $2^{|X_t|}3^{|X_t|^2}2^{|X_t|^2}\leq 3^{2|X_t|^2}$ configurations for a bag $X_t$. Hence, in total the running time is upper-bounded by $2^{O(b^2)}n$, where $b$ is the maximum size of a bag in $(\mathscr{P},\mathcal{X})$. Since any bag induces a subgraph of $G$ of diameter at most~$4$, by Lemma~\ref{lemma:bnded-diameter-subgraph}, $b=O(k^2)$. Therefore $2^{O(b^2)}n=2^{O(k^4)}n$, as claimed.
\end{proof}

\section{Conclusion}\label{sec:conclu}

We proved that \PBLD, \PBOLD, \PBID{} and \PBMD{} are NP-complete even for interval graphs that have diameter~$2$ and for permutation graphs that have diameter~$2$.
 This is in contrast to related problems such as \textsc{Dominating Set}, which is linear-time solvable both on interval graphs and on permutation graphs. However, we do not know their complexity for unit interval graphs or bipartite permutation graphs. Note that both \PBLD{} and \PBMD{} are polynomial-time solvable on chain graphs, a subclass of bipartite permutation graphs~\cite{FHvMS15}. Probably the same approach as in~\cite{FHvMS15} would also work for \PBOLD{} and \PBID.

Contrary to what we claimed in the conference version of this paper~\cite{WGversion}, our reduction gadgets are not interval graphs and permutation graphs at the same time. Hence, we leave it as an open question to determine the complexity of the studied problems when restricted to graphs which are both interval and permutation graphs. Similarly, it could be interesting to determine their complexity for graphs that are both split graphs and interval graphs, or split graphs and permutation graphs.

We remark that our generic reduction would also apply to related
problems that have been considered in the literature, such as
\textsc{Locating-Total Dominating Set}~\cite{hr12} or
\textsc{Differentiating-Total Dominating Set}~\cite{C08}.

Regarding our positive result that \PBMD{} parameterized by the solution size is FPT on interval graphs, an interesting question is whether it can be extended to other graph classes, such as permutation graphs. Another interesting class is the one of chordal graph, since it is a proper superclass of both interval graphs and split graphs, both of which admit an FPT algorithm for \PBMD. During the revision of this paper, it was brought to our knowledge that in a recent paper, Belmonte, Fomin, Golovach and Ramanujan~\cite{BFGR15} have answered these questions by showing that for any class of graphs of bounded tree-length, \PBMD{} is FPT when parameterized by the solution size. Examples of such classes are the ones of chordal graphs, asteroidal triple-free graphs and permutation graphs.

\vspace{0.2cm}

\subsection*{Acknowledgements}
We thank Adrian Kosowski for helpful preliminary discussions on the topic of this paper. We are also grateful to the reviewers for their useful comments which subsequently made the paper clearer.

\end{document}